\documentclass[a4paper]{article}

\usepackage{amsmath}
\usepackage{amssymb}
\usepackage{amsfonts}
\usepackage{amsthm}
\usepackage{ascmac}
\usepackage{bm}
\usepackage{color}
\usepackage{mathrsfs}
\usepackage{threeparttable}
\usepackage{graphicx}
\usepackage{multirow}
\usepackage{multicol}
\usepackage{comment}

\newcommand{\e}{\mathrm{e}}

\renewcommand{\i}{\mathrm{i}}
\renewcommand{\ss}{\mathrm{ss}}

\newcommand{\const}{\mathrm{const.}}
\newcommand{\A}{\mathrm{A}}
\newcommand{\B}{\mathrm{B}}
\newcommand{\C}{\mathrm{C}}
\newcommand{\D}{\mathrm{D}}

\newcommand{\I}{\mathrm{I}}
\renewcommand{\S}{\mathrm{S}}
\newcommand{\R}{\mathcal{R}}
\newcommand{\RR}{\mathrm{R}}

\newcommand{\ep}{\varepsilon}
\renewcommand{\Pi}{\varPi}

\title{The SIS Competition Model for Conflicting Rumors}
\author{Yu Takiguchi and Koji Nemoto}

\begin{document}

\maketitle

\section*{Abstract}

We propose an SIS competition model describing the propagation of conflicting rumors, such as fake news and its corrections. 
This simple model captures the interaction between rumor propagation and opinion dynamics, where rumors drive opinion changes and, conversely, individuals' opinions determine the infection rates of rumors. 
We analytically derive all steady states and their stability. 
These results uncover a novel coexistence mechanism. 
This coexistence corresponds to a scenario where belief in one rumor (e.g., fake news) paradoxically aids the spread of the opposing rumor (e.g., corrective information). 
Due to this mechanism, a nontrivial but realistic phenomenon occurs where a lower infection rate actually enhances the spread of a rumor. 
Furthermore, although the model does not explicitly incorporate majority conformity, a phenomenon where the majority gains an advantage emerges spontaneously. 
Consequently, even if one rumor has a higher infection rate, it may be eliminated by the other if its initial share fails to exceed a critical threshold. 
We analytically derive this threshold using the singular perturbation method.

\section{Introduction}

The phenomenon of conflicting rumors spreading simultaneously, such as fake news versus its corrections, 
support versus criticism of a specific politician, or conspiracy theories versus anti-conspiracy arguments, is frequently observed. 
Under what conditions can fake news be eradicated by corrective information? 
What mechanisms allow conflicting rumors to coexist? 
Addressing these questions is crucial in today's society characterized by developed social media.

Theoretical studies on rumor propagation have been advanced using epidemic models\cite{deArruda2016, Caldarelli2025,Pastor-Satorras2015}. 
As early as 1964, Daley and Kendall proposed the SIR model (DK model), consisting of individuals who are ignorant of the rumor (susceptible, $\S$), 
spreading the rumor (infected, $\I$), or have ceased spreading it (recovered/removed, $\RR$)\cite{Daley1964}.
As a model describing multiple rumors, the 2SI2R model has been proposed\cite{Wang2014, Zhang2018}. 
This model assumes a situation where an old rumor $\A$ and a new rumor $\B$ propagate. 
Individuals can take five states: ignorant of both rumors ($\S$), spreading either rumor ($\I_\A$, $\I_\B$), 
or knowing but not spreading either rumor ($\RR_\A$, $\RR_\B$). 
Those who know the old rumor $\A$ ($\I_\A$, $\RR_\A$) can transition to state $\I_\B$, but transitions from the new rumor to the old one are not considered. 
In this model, the threshold for global rumor diffusion (i.e., a pandemic) has been derived.
Karrer and Newman\cite{Karrer2011} also studied an SIR model for two interacting diseases, 
showing that even diseases with slow spreading speeds can cause a pandemic if their transmission probability is high.
Trpevski et al.\cite{Trpevski2010} extended the SIS model to investigate the diffusion of two rumors by considering three states: $\S$, $\I_\A$, and $\I_\B$. 
In their model, assuming one rumor is more credible or reputable, they assigned asymmetric transition rules for $\A$ and $\B$. 
They revealed through simulation that two rumors can coexist if the social network meets certain conditions regarding properties such as clustering and degree.
The situation where two types of diseases spread on respective transmission networks has been investigated by Sanz et al.\cite{Sanz2014}.
Additionally, Chen et al.\cite{Chen2022} have comprehensively reviewed epidemic models describing multiple rumors.

Which rumor spreads more easily depends on people's values and opinions.
For example, ardent supporters of a political party will actively spread positive rumors about that party, while ignoring or attacking negative ones.
Recently, the "echo chamber phenomenon", where people with similar values connect densely, amplifying rumors aligned with their values while excluding critical ones, has become a growing concern\cite{Roozenbeek2024}.
However, previous studies using epidemic models\cite{Wang2014,Zhang2018,Karrer2011,Trpevski2010,Sanz2014,Chen2022} have not represented the aspect where rumor propagation is influenced by each individual's values. 

The dynamics of how opinions change through communication have been studied using opinion models\cite{Caldarelli2025}.
In the voter model\cite{Liggett1985,Liggett1999,Fernandez2014}, the simplest opinion model, a randomly selected individual chooses a neighbor at random and adopts the neighbor's opinion as their own.
Conditions under which only one opinion eventually survives, that is, reaching a consensus, have been actively studied. 
Recently, more concrete opinion models motivated by social media have been proposed\cite{Baumann2020,Hata2025,Sasahara2021}.

Epidemic models describe the propagation of rumors, while opinion models describe the evolution of beliefs.
In real society, people change their opinions under the influence of rumors they hear, and the transmissibility of rumors changes under the influence of the opinions people support.
Therefore, to analyze the competitive diffusion of conflicting rumors, it is necessary to integrate these two models. 
Thus, we propose an SIS competition model that describes both rumor propagation and opinion dynamics.
Using this model, we analytically clarify the conditions under which only one rumor survives.
Furthermore, we uncover a previously unreported mechanism that enables the coexistence of rumors.

\section{Model}

\begin{figure}[hbtp]
    \centering
    \includegraphics[width=0.3\linewidth]{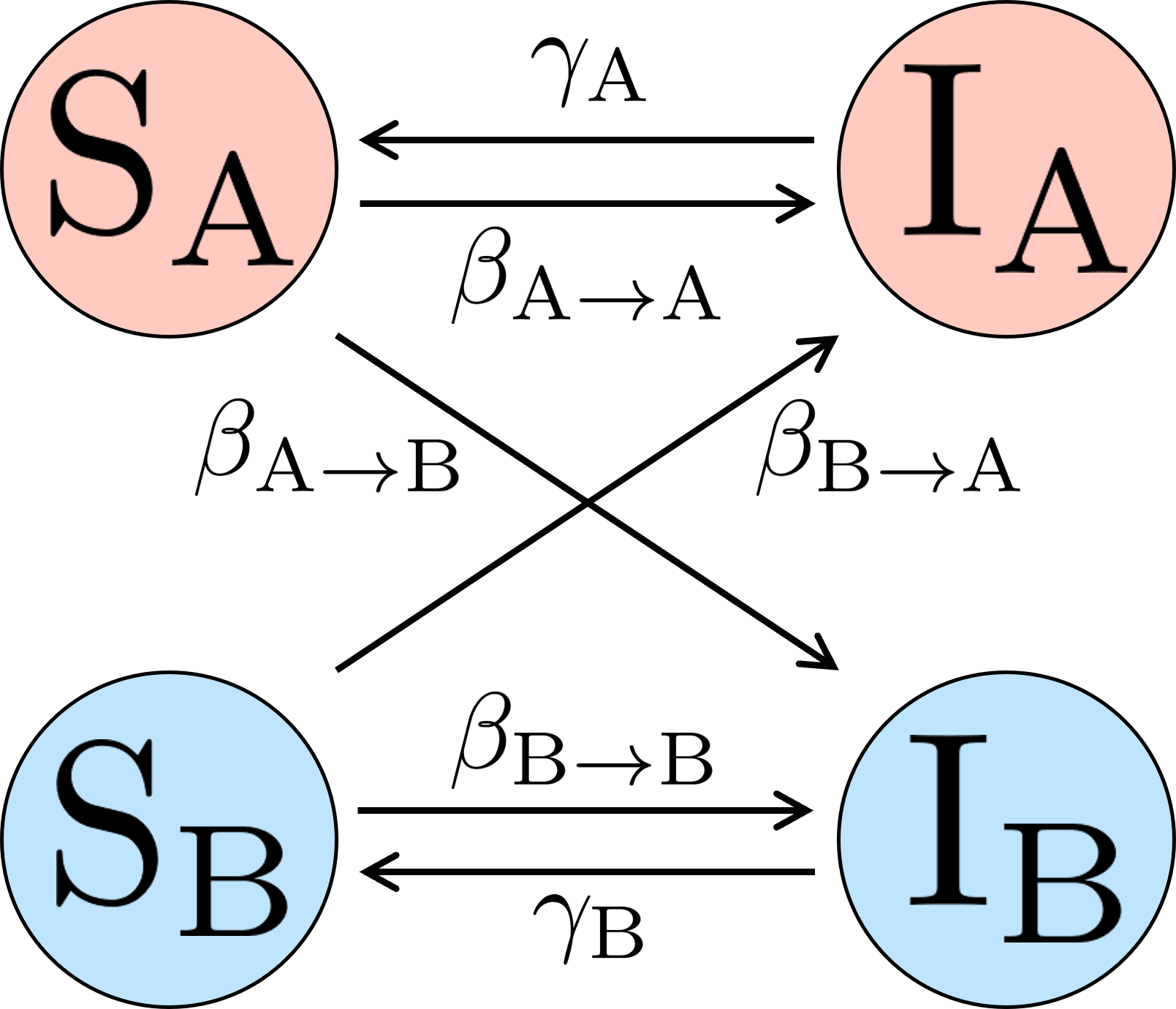}
    \caption{
        Schematic of transitions in the SIS competition model.
        \label{Fig_compartment}
    }
\end{figure}

We construct a model describing the propagation of rumors regarding two exclusive opinions $\A$ and $\B$ by extending the SIS model.
Let $\rho_r$ be the fraction of individuals who support opinion $r \in \{\A,\B\}$.
For simplicity, we assume that every individual supports one of the opinions, i.e., $\rho_{\A} + \rho_{\B} = 1$ holds.

Each individual takes either state $\S_r$ ($r \in \{\A, \B\}$), representing those who support opinion $r$ but do not spread the rumor, 
or state $\I_r$, representing those who support opinion $r$ and are spreading the rumor.
Denoting the fractions of individuals in these states as $S_r$ and $I_r$, the relation $\rho_r = S_r + I_r$ holds.
An individual in state $\S_{r'}$ hears the rumor promoting $r$ from an individual in state $\I_r$ and starts spreading $r$ at a transmission rate $\beta_{r' \to r}$.
The case $r \neq r'$ corresponds to a change of opinion triggered by hearing the rumor.
An individual spreading a rumor $r$ transitions to state $\S_r$ at a recovery rate $\gamma_r$.
This implies that they lose interest and cease spreading the rumor.
Fig.~\ref{Fig_compartment} summarizes these transitions.
This model, which we call the SIS competition model, is described by the following ordinary differential equations:
\begin{align}
    \dot{S}_{\A} &= - \beta_{\A \to \A} S_{\A} I_{\A} - \beta_{\A \to \B} S_{\A} I_{\B} + \gamma_{\A} I_{\A}, \label{Eq_S_A}\\  
    \dot{I}_{\A} &= + \beta_{\A \to \A} S_{\A} I_{\A} + \beta_{\B \to \A} S_{\B} I_{\A} - \gamma_{\A} I_{\A}, \label{Eq_I_A}\\  
    \dot{S}_{\B} &= - \beta_{\B \to \B} S_{\B} I_{\B} - \beta_{\B \to \A} S_{\B} I_{\A} + \gamma_{\B} I_{\B}, \label{Eq_S_B}\\  
    \dot{I}_{\B} &= + \beta_{\B \to \B} S_{\B} I_{\B} + \beta_{\A \to \B} S_{\A} I_{\B} - \gamma_{\B} I_{\B}. \label{Eq_I_B}
\end{align}
The transition from $\S$ to $\I$ represents the propagation of a rumor, while the transition from $r$ to $r'$ represents a change in an individual's opinion.
Thus, our model describes both rumor propagation and opinion dynamics.
Note that in this model, we assume that a change of opinion results in a transition from $\S_{r}$ to $\I_{r'}$.
This assumption is based on the idea that if one hears a rumor impactful enough to alter their opinion, they will likely be motivated to spread that rumor to others.
Although a term for the transition from $\S_{r}$ to $\S_{r'}$ could be added, we have confirmed that it does not lead to qualitative differences in the results.

We define the basic reproduction numbers $\R_{0}^{\A}$, $\R_{0}^{\B}$, $\R_{0}^{\B \to \A}$, and $\R_{0}^{\A \to \B}$ of the SIS competition model as follows:
\begin{align}
    \R_{0}^{\A} &\equiv \frac{\beta_{\A \to \A} }{\gamma_{\A}},
    &\R_{0}^{\B} &\equiv \frac{\beta_{\B \to \B} }{\gamma_{\B}},
    &\R_{0}^{\B \to \A} &\equiv \frac{\beta_{\B \to \A} }{\gamma_{\A}},
    &\R_{0}^{\A \to \B} &\equiv \frac{\beta_{\A \to \B} }{\gamma_{\B}}.
\end{align}
The basic reproduction number $\R_{0}^{r}$ represents the number of individuals in state $\I_r$ (i.e., secondary infections) generated by a single individual in state $\I_r$, in a population where everyone else is in state $\S_r$.
Similarly, $\R_{0}^{r' \to r}$ represents the number of individuals in state $\I_{r}$ generated by a single individual in state $\I_{r}$, in a population where everyone else is in state $\S_{r'}$.
In this model, four basic reproduction numbers are defined according to the combination of the currently supported opinion and the newly adopted opinion.
We define the effective reproduction numbers $\R_{t}^{\A}(S_{\A}, S_{\B})$ and $\R_{t}^{\B}(S_{\A}, S_{\B})$ as follows:
\begin{align}
    \R_{t}^{\A}(S_{\A}, S_{\B})
    &\equiv 
    \R_{0}^{\A} S_{\A} + \R_{0}^{\B \to \A} S_{\B},\\
    \R_{t}^{\B}(S_{\A}, S_{\B})
    &\equiv 
    \R_{0}^{\B} S_{\B} + \R_{0}^{\A \to \B} S_{\A}.
\end{align}
The effective reproduction number $\R_{t}^{r}$ represents the number of individuals in state $\I_{r}$ generated by an individual in state $\I_{r}$ before they recover.
$\R_{t}^{\A}$ can be expressed as the sum of $\R_{0}^{\A} S_{\A}$, the number of individuals changing state from $\S_{\A}$ to $\I_{\A}$, and $\R_{0}^{\B \to \A} S_{\B}$, the number of individuals changing state from $\S_{\B}$ to $\I_{\A}$.

\section{Results}

\subsection{Linear Stability Analysis}

We obtain the steady states by setting $\dot{S}_{\A} = \dot{I}_{\A} = \dot{S}_{\B} = \dot{I}_{\B} = 0$.
From Eq.~\eqref{Eq_I_A}, we obtain $I_{\A} = 0$ or $\beta_{\A \to \A} S_{\A}  + \beta_{\B \to \A} S_{\B} - \gamma_{\A} = 0$.
The former corresponds to a steady state where spreaders of rumor $\A$ die out, while the latter corresponds to a steady state where spreaders of rumor $\A$ survive.
Similarly, Eq.~\eqref{Eq_I_B} yields $I_{\B} = 0$ or $\beta_{\B \to \B} S_{\B}  + \beta_{\A \to \B} S_{\A} - \gamma_{\B} = 0$.
The system admits four steady states, corresponding to the combinations of these conditions.

\subsubsection{Steady state where neither rumor spreads}

First, we consider the steady state where neither rumor promoting opinion $\A$ nor $\B$ is spread.
Setting $I_{\A} = I_{\B} = 0$ implies $\dot{S}_{\A} = \dot{S}_{\B} = 0$ from Eqs. \eqref{Eq_S_A} and \eqref{Eq_S_B}. 
Thus, the steady state is given by
\begin{align}
    (S_{\A}, I_{\A}, S_{\B}, I_{\B}) = (\rho_{\A}, 0, 1 - \rho_{\A}, 0)
    \qquad 0 \leq \rho_{\A} \leq 1.
\end{align}
Note that this steady state forms a line segment in the $(S_{\A}, I_{\A}, S_{\B}, I_{\B})$ phase space, rather than a single point.

The eigenvalues $\lambda$ of the Jacobian matrix are
\begin{align}
    \lambda 
    &= 0,\ \beta_{\A \to \A} \rho_{\A} + \beta_{\B \to \A} \rho_{\B} - \gamma_{\A},\ \beta_{\B \to \B} \rho_{\B} + \beta_{\A \to \B} \rho_{\A} - \gamma_{\B}\nonumber\\
    &= 0,\ \gamma_{\A}  \left( \R_{t}^{\A} (\rho_{\A}, \rho_{\B}) - 1 \right), \gamma_{\B}  \left( \R_{t}^{\B} (\rho_{\A}, \rho_{\B}) - 1 \right).
\end{align}
The eigenvalue $\lambda = 0$ is doubly degenerate, and its eigenvectors correspond to directions in which $\rho_{\A}$ varies while maintaining $I_{\A} = I_{\B} = 0$.
Since the state $(S_{\A}, I_{\A}, S_{\B}, I_{\B}) = (\rho_{\A}, 0, 1 - \rho_{\A}, 0)$ is stationary for any arbitrary $\rho_{\A}$, a perturbation in the direction of these eigenvectors results in neither convergence to nor divergence from the original steady state.
The remaining two eigenvalues are both negative when $\R_{t}^{\A} < 1$ and $\R_{t}^{\B} < 1$.
Under these conditions, this steady state is locally stable.
Recall that $\R_{t}^{r}$ represents the number of individuals in state $\I_{r}$ generated by a single individual in state $\I_{r}$.
In other words, this steady state is locally stable if the number of secondary infections remains below $1$ when an individual in state $\I_{\A}$ or $\I_{\B}$ appears.
In the standard SIS model, the stability of the disease-free equilibrium is determined by the basic reproduction number, whereas in this model, it is determined by the effective reproduction number.
This stems from the fact that while $S \approx 1$ when $I \ll 1$ in the standard SIS model, $S_r \approx \rho_r$ when $I_{\A}, I_{\B} \ll 1$ in our model.

We determine the region in the parameter space $(\R_{0}^{\A}, \R_{0}^{\B}, \R_{0}^{\B \to \A}, \R_{0}^{\A \to \B} )$ where this steady state exists and is locally stable.
This corresponds to finding the parameter region where a value of $\rho_{\A}$ satisfying $\R_{t}^{\A}(\rho_{\A}, \rho_{\B}) < 1$ and $\R_{t}^{\B}(\rho_{\A}, \rho_{\B}) < 1$ exists within the range $0 \leq \rho_{\A} \leq 1$.
Solving for this condition yields the boundary line defined by
\begin{align}
(\R_{0}^{\A} - 1)(\R_{0}^{\B} - 1) = (\R_{0}^{\B \to \A} - 1) (\R_{0}^{\A \to \B} - 1).
\end{align}

\subsubsection{Steady state where only one rumor spreads}

Next, we consider the steady state where only the rumor promoting opinion $\A$ is spreading in the population.
In this case, since $I_{\B} = 0$, $I_{\A} > 0$, and the condition $\beta_{\A \to \A} S_{\A}  + \beta_{\B \to \A} S_{\B} - \gamma_{\A} = 0$ holds, the steady state is given by
\begin{align}
    (S_{\A}, I_{\A}, S_{\B}, I_{\B}) = \left(\frac{1}{\R_{0}^{\A}}, 1 - \frac{1}{\R_{0}^{\A}}, 0, 0 \right) \equiv (S_{\A}^{\ss}, I_{\A}^{\ss}, 0, 0).
\end{align}
The steady state where only one rumor spreads is identical to the steady state of the standard SIS model.

The eigenvalues $\lambda$ of the Jacobian matrix are
\begin{align}
    \lambda 
    &= 0,\ - \beta_{\A \to \A} I_{\A}^{\ss},\ - \beta_{\B \to \A} I_{\A}^{\ss},\ \beta_{\A \to \B} S_{\A}^{\ss} - \gamma_{\B}.
\end{align}
The zero eigenvalue stems from the normalization condition and does not affect the stability of the system.
The second and third eigenvalues become negative when $\R_{0}^{\A} > 1$.
This implies that in a population where everyone is in state $\S_{\A}$, the rumor $\A$ propagates globally; in other words, a pandemic occurs.
The fourth eigenvalue becomes negative when $\R_{0}^{\A} >  \R_{0}^{\A \to \B}$.
This implies that in a population where everyone is in state $\S_{\A}$, the rumor $\A$ spreads more easily than the rumor $\B$.

A similar argument holds for the steady state where only the rumor $\B$ spreads.
The steady state
\begin{align}
(S_{\A}, I_{\A}, S_{\B}, I_{\B}) = \left(0, 0, \frac{1}{\R_{0}^{\B}}, 1 - \frac{1}{\R_{0}^{\B}}\right) \equiv (0, 0, S_{\B}^{\ss}, I_{\B}^{\ss})
\end{align}
is locally stable when $\R_{0}^{\B} > 1$ and $\R_{0}^{\B} > \R_{0}^{\B \to \A}$.

\subsubsection{Steady state where both rumors spread}

Finally, the steady state where both rumors promoting opinions $\A$ and $\B$ are spreading in the population is given by
\begin{align}
    (S_{\A}, I_{\A}, S_{\B}, I_{\B}) &=  (S_{\A}^{*\ss}, I_{\A}^{*\ss}, S_{\B}^{*\ss}, I_{\B}^{*\ss}),\\
    S_{\A}^{*\ss} &\equiv \frac{\R_{0}^{\B \to \A} - \R_{0}^{\B}}{\R_{0}^{\B \to \A} \R_{0}^{\A \to \B} - \R_{0}^{\A}\R_{0}^{\B}},\\
    S_{\B}^{*\ss} &\equiv \frac{\R_{0}^{\A \to \B}- \R_{0}^{\A}}{\R_{0}^{\B \to \A} \R_{0}^{\A \to \B} - \R_{0}^{\A}\R_{0}^{\B}},\\
    I_{\A}^{*\ss} 
    &\equiv \frac{\beta_{\A \to \B} S_{\A}^{*\ss} }{\beta_{\A \to \B} S_{\A}^{*\ss} + \beta_{\B \to \A} S_{\B}^{*\ss}} (1 - S_{\A}^{*\ss} - S_{\B}^{*\ss}),\\
    I_{\B}^{*\ss} 
    &\equiv \frac{\beta_{\B \to \A} S_{\B}^{*\ss}}{\beta_{\A \to \B} S_{\A}^{*\ss} + \beta_{\B \to \A} S_{\B}^{*\ss}} (1 - S_{\A}^{*\ss} - S_{\B}^{*\ss}).
\end{align}
We determine the parameter region where this steady state satisfies $0 \leq S_{\A}^{*\ss}$, $I_{\A}^{*\ss}$, $S_{\B}^{*\ss}$, $I_{\B}^{*\ss} \leq 1$.
The conditions $0 \leq S_{\A}^{*\ss}$ and $0 \leq S_{\B}^{*\ss}$ are satisfied when the signs of $\R_{0}^{\A \to \B} - \R_{0}^{\A}$ and $\R_{0}^{\B \to \A} - \R_{0}^{\B}$ match.
The parameter region satisfying $S_{\A}^{*\ss} + S_{\B}^{*\ss} \leq 1$ is given by
\begin{align}
    (\R_{0}^{\A} - 1) (\R_{0}^{\B} - 1)
    &\leq
    (\R_{0}^{\B \to \A} - 1) (\R_{0}^{\A \to \B} - 1)
\end{align}
when $\R_{0}^{\A \to \B} > \R_{0}^{\A}$ and $\R_{0}^{\B \to \A} > \R_{0}^{\B}$, and by
\begin{align}
    (\R_{0}^{\A} - 1) (\R_{0}^{\B} - 1)
    &\geq
    (\R_{0}^{\B \to \A} - 1) (\R_{0}^{\A \to \B} - 1)
\end{align}
when $\R_{0}^{\A \to \B} < \R_{0}^{\A}$ and $\R_{0}^{\B \to \A} < \R_{0}^{\B}$.
When the conditions $0 \leq S_{\A}^{*\ss}, S_{\B}^{*\ss}$ and $S_{\A}^{*\ss} + S_{\B}^{*\ss} \leq 1$ are satisfied, $0 \leq I_{\A}^{*\ss} \leq 1$ and $0 \leq I_{\B}^{*\ss} \leq 1$ always hold.

Linear stability analysis reveals that the steady state where rumors coexist becomes locally stable when $\R_{0}^{\A \to \B} > \R_{0}^{\A}$ and $\R_{0}^{\B \to \A} > \R_{0}^{\B}$ (see Appendix~\ref{Sec_Hurwitz}).
This condition implies that individuals are more likely to spread a rumor promoting an opinion they did not previously support (accompanied by a change of opinion) 
than to spread a rumor promoting their currently supported opinion.

\subsubsection{Phase Diagram}

\begin{figure}[hbtp]
    \centering
    \includegraphics[width=0.6\linewidth]{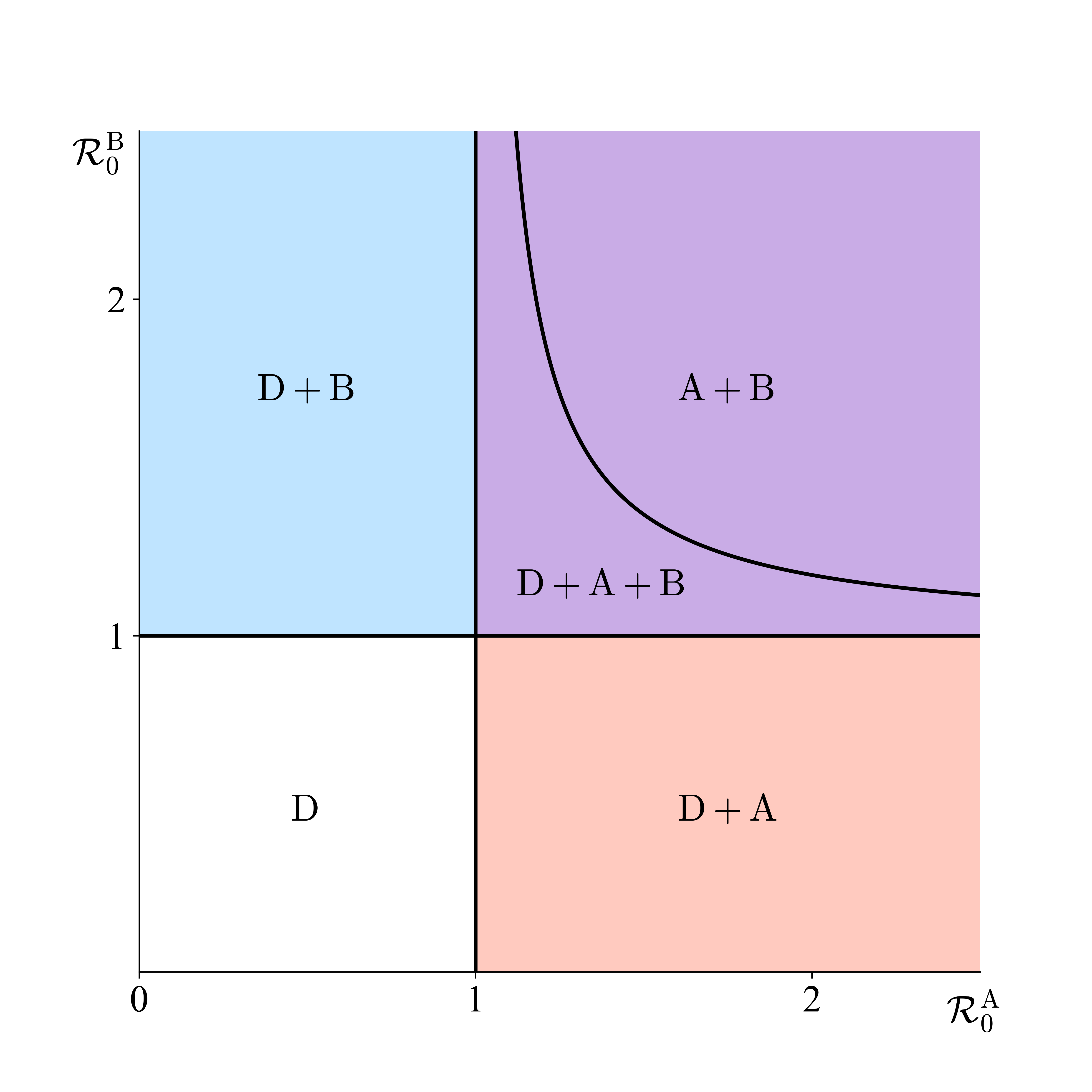}
    \caption{
        Phase diagram of the SIS competition model for $\R_{0}^{\B \to \A}, \R_{0}^{\A \to \B} < 1$.
        The horizontal and vertical axes represent $\R_{0}^{\A}$ and $\R_{0}^{\B}$, respectively.
        For example, $\D + \A$ indicates that the system is in either the $\D$ phase or the $\A$ phase.
        Parameters: $\R_{0}^{\B \to \A}=0.1, \R_{0}^{\A \to \B}=0.8$.
        The curve is given by $(\R_{0}^{\A} - 1) (\R_{0}^{\B} - 1) = (\R_{0}^{\B \to \A} - 1) (\R_{0}^{\A \to \B} - 1)$.
        \label{Fig_phase_diagram_1}
    }
\end{figure}

\begin{figure}[hbtp]
    \centering
    \includegraphics[width=0.6\linewidth]{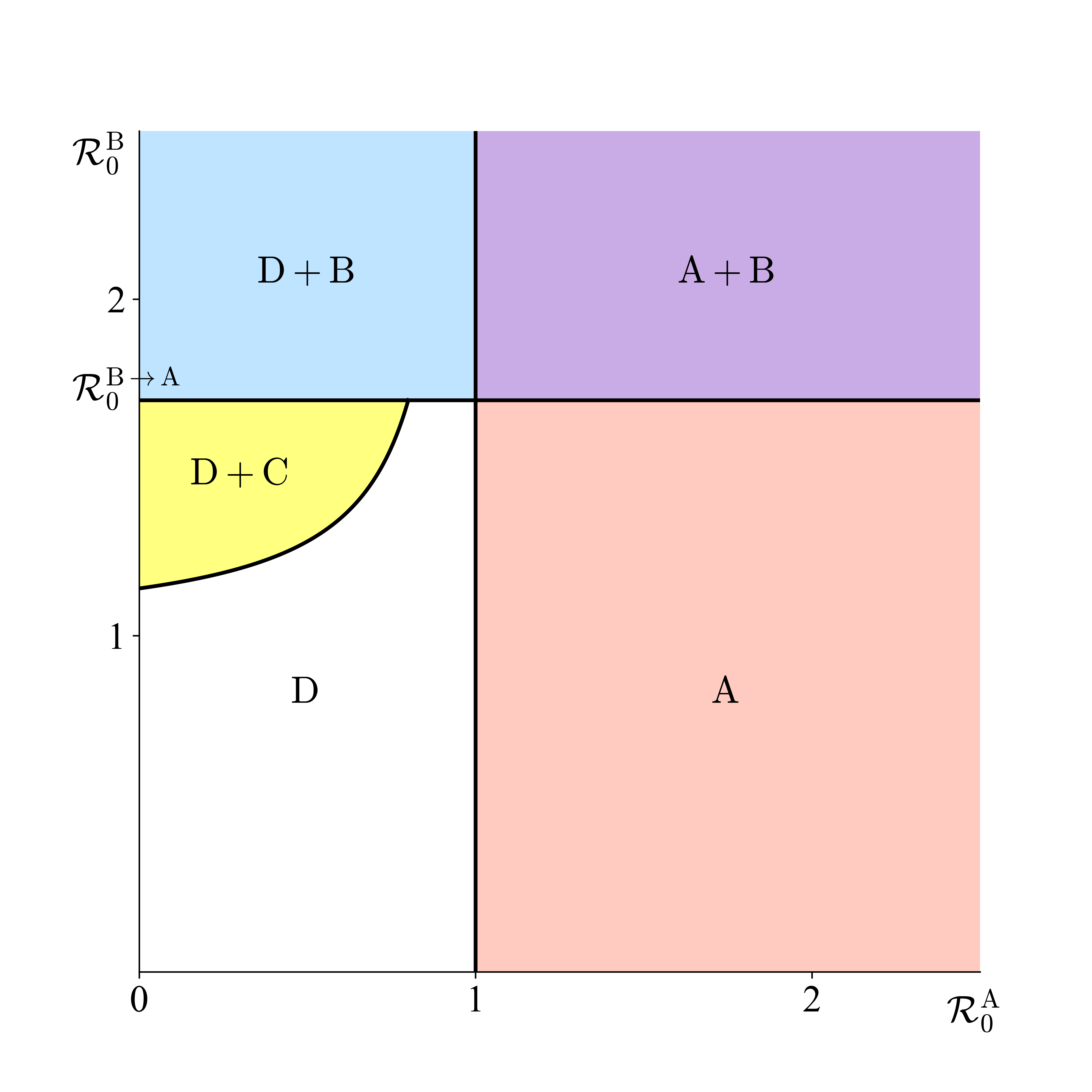}
    \caption{
        Phase diagram of the SIS competition model for $\R_{0}^{\A \to \B} < 1 < \R_{0}^{\B \to \A}$.
        The horizontal and vertical axes represent $\R_{0}^{\A}$ and $\R_{0}^{\B}$, respectively.
        Parameters: $\R_{0}^{\B \to \A}=1.7, \R_{0}^{\A \to \B}=0.8$.
        The curve is given by $(\R_{0}^{\A} - 1) (\R_{0}^{\B} - 1) = (\R_{0}^{\B \to \A} - 1) (\R_{0}^{\A \to \B} - 1)$.
        \label{Fig_phase_diagram_2}
    }
\end{figure}

\begin{figure}[hbtp]
    \centering
    \includegraphics[width=0.6\linewidth]{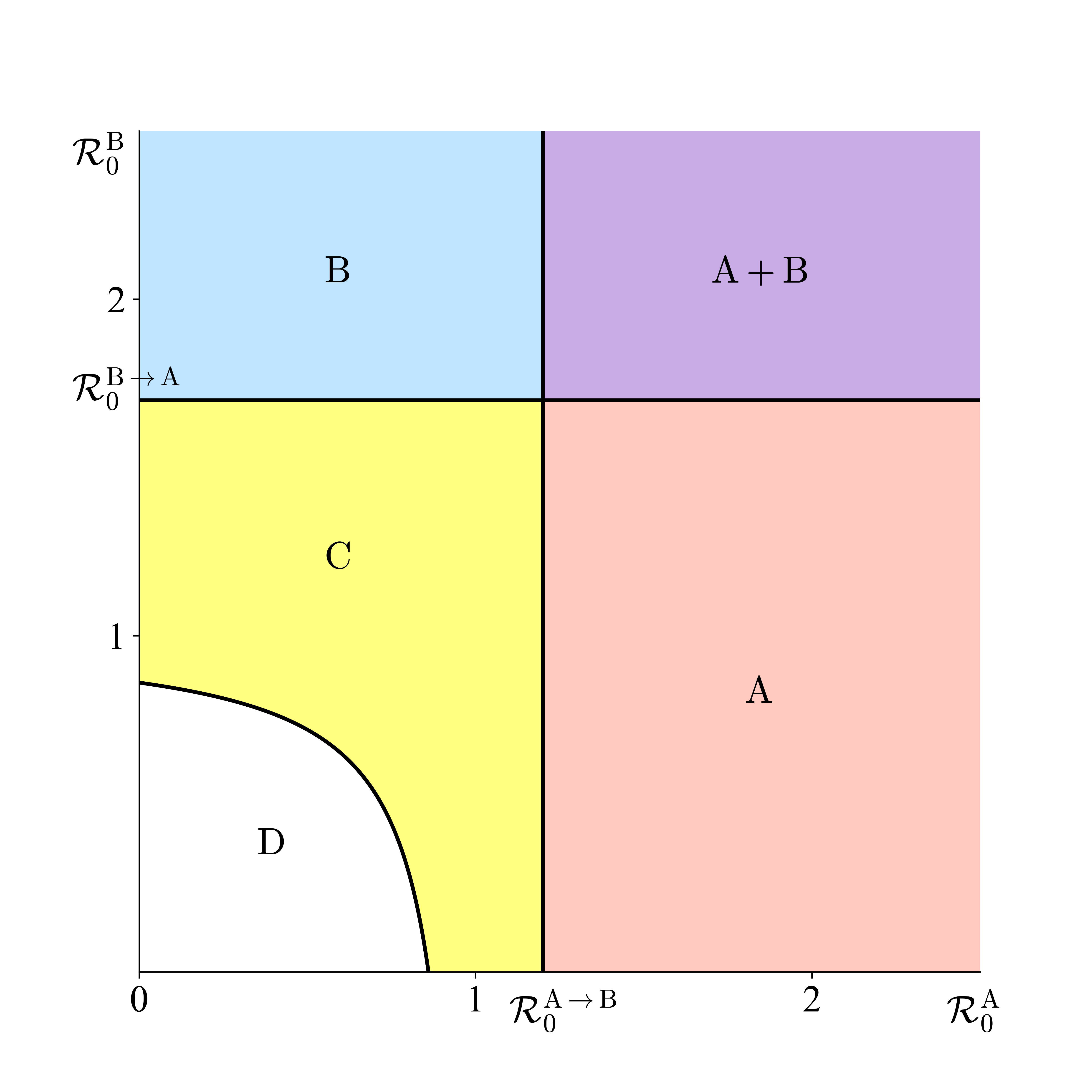}
    \caption{
        Phase diagram of the SIS competition model for $1 < \R_{0}^{\B \to \A}, \R_{0}^{\A \to \B}$.
        The horizontal and vertical axes represent $\R_{0}^{\A}$ and $\R_{0}^{\B}$, respectively.
        Parameters: $\R_{0}^{\B \to \A}=1.7, \R_{0}^{\A \to \B}=1.2$.
        The curve is given by $(\R_{0}^{\A} - 1) (\R_{0}^{\B} - 1) = (\R_{0}^{\B \to \A} - 1) (\R_{0}^{\A \to \B} - 1)$.
        \label{Fig_phase_diagram_3}
    }
\end{figure}

\begin{figure}[hbtp]
    \centering
    \includegraphics[width=0.6\linewidth]{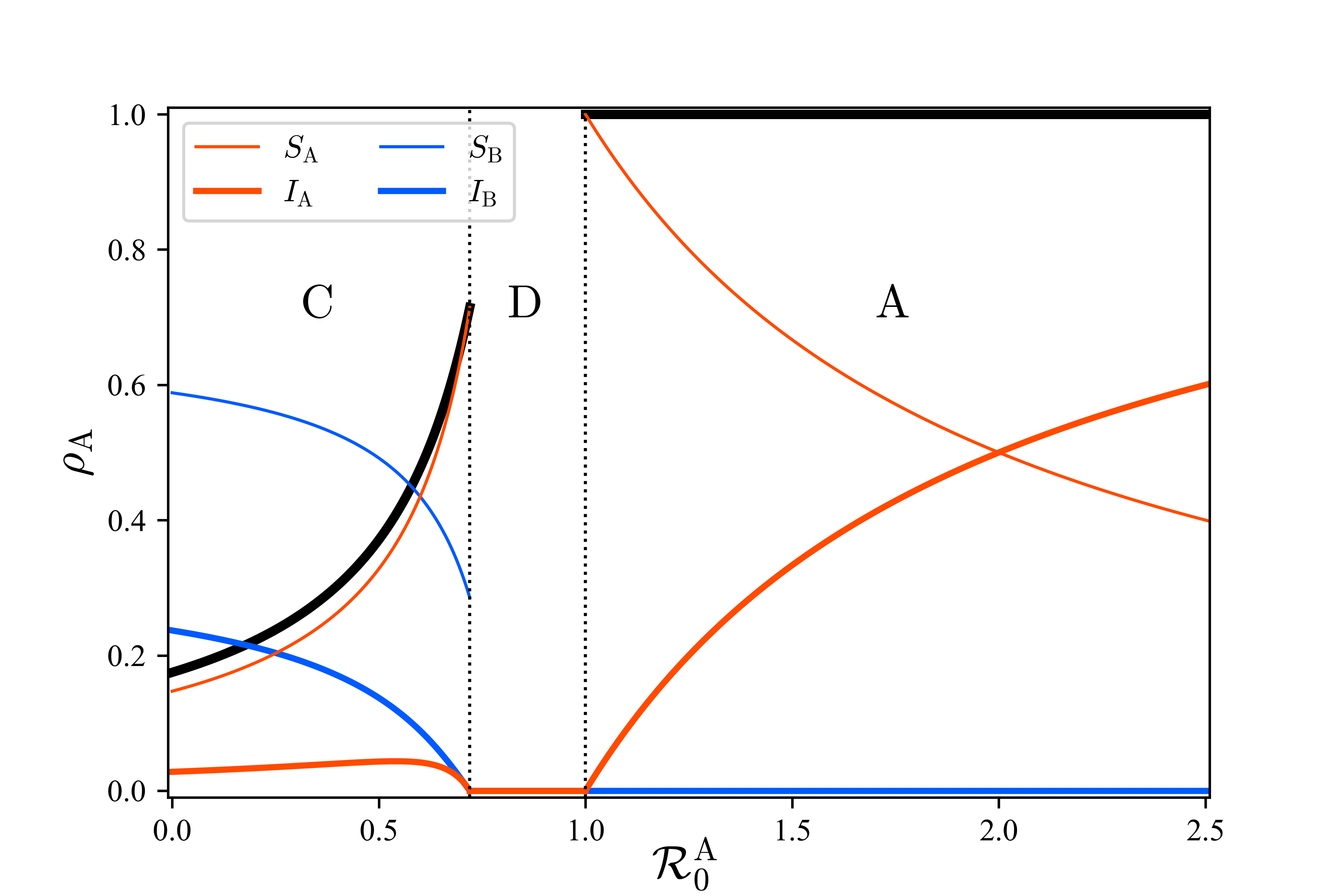}
    \caption{
        Steady state at $\R_{0}^{\B} = 1.5$ in Fig.~\ref{Fig_phase_diagram_2}.
        The horizontal axis represents the basic reproduction number $\R_{0}^{\B}$, and the vertical axis represents the fraction of each state.
        The red and blue lines represent the fractions of individuals supporting opinions $\A$ and $\B$, respectively, while the black solid line represents $\rho_{\A}$.
        The steady states in the $\D$ phase are partially omitted.
        The boundary of the $\C$ phase is given by $\R_{0}^{\A} = 1 - (1 - \R_{0}^{\A \to \B} ) (\R_{0}^{\B \to \A} - 1)(\R_{0}^{\B} - 1)^{-1}$.
        \label{Fig_reentrant}  
    }
\end{figure}

In the parameter space $(\R_{0}^{\A}, \R_{0}^{\B}, \R_{0}^{\B \to \A}, \R_{0}^{\A \to \B} )$, 
we define the parameter regions where the steady state in which only rumor $\A$ or $\B$ spreads is locally stable as the $\A$ phase and the $\B$ phase, respectively.
The region where the steady state in which both rumors coexist is locally stable is defined as the $\C$ phase, 
and the region where the steady state in which both rumors die out is locally stable is defined as the $\D$ phase.
Note that while there are originally six parameters $\beta_{r\to r'}$ and $\gamma_r$ ($r,r' \in \{ \A,\B \}$), there are only four basic reproduction numbers.
Since the local stability of each steady state depends solely on the basic reproduction numbers, the phases are completely specified by these four parameters.
However, since $I_{\A}^{*\ss}$ and $I_{\B}^{*\ss}$ depend on the explicit values of $\beta_{r \to r'}$, the values of the steady state where rumors coexist are not determined solely by the basic reproduction numbers.
The phases defined above are not mutually exclusive.
For example, a parameter region where $\R_{0}^{\A} > \R_{0}^{\A \to \B} > 1$ and $\R_{0}^{\B} > \R_{0}^{\B \to \A} > 1$ corresponds to either the $\A$ phase or the $\B$ phase.
This means that eventually only rumor $\A$ or $\B$ will survive, and which steady state is realized depends on the initial conditions.

The phase diagrams of the SIS competition model are shown in Fig.~\ref{Fig_phase_diagram_1}--\ref{Fig_phase_diagram_3}.
Fig.~\ref{Fig_phase_diagram_1} shows the case where both basic reproduction numbers related to opinion change are less than $1$ ($\R_{0}^{\B \to \A}, \R_{0}^{\A \to \B} < 1$).
The boundary of the $\A$ phase is $\R_{0}^{\A} = 1$, which is the same as the pandemic threshold in the standard SIS model, and similarly, the boundary of the $\B$ phase is $\R_{0}^{\B} = 1$.
The existence of the $\D$ phase even when $\R_{0}^{\A} > 1$ and $\R_{0}^{\B} > 1$ is due to the fact that since opinion change is unlikely, the diffusion of rumor $r$ is inhibited by individuals in states $\S_{r'}$ and $\I_{r'}$ ($r'\neq r$).
In this parameter region, rumors cannot coexist.

Fig.~\ref{Fig_phase_diagram_2} shows the case where changing opinion from $\B$ to $\A$ is easy ($\R_{0}^{\B \to \A} > 1$), but changing from $\A$ to $\B$ is difficult ($\R_{0}^{\A \to \B} < 1$).
In this parameter region, rumors can coexist.
The separation between the $\C$ phase and the $\A$ phase is a nontrivial result.
Fig.~\ref{Fig_reentrant} depicts the steady state in this parameter region.
As $\R_{0}^{\A}$ is gradually increased from $0$, rumor $\A$ initially survives in the steady state, but becomes extinct beyond a certain point. 
As $\R_{0}^{\A}$ is further increased beyond $1$, the system enters the $\A$ phase, allowing rumor $\A$ to survive again.
Since the $\C$ phase lies in the region $\R_{0}^{\A} < 1$, the supporters of opinion $\A$ alone cannot spread the rumor globally.
On the other hand, since $\R_{0}^{\B \to \A} > 1$, if there are many individuals in state $\S_{\B}$ in the population, rumor $\A$ can be spread by causing them to change their opinion.
In other words, if a rumor regarding $\A$ is trivial to its supporters ($\R_{0}^{\A} < 1$) 
but shocking to supporters of opinion $\B$ ($\R_{0}^{\B \to \A} > 1$), 
yet remains insufficient to eradicate opinion $\B$, these conflicting rumors can coexist.
If $\R_{0}^{\A}$ is large but remains less than $1$, $S_{\B}$ decreases, which ironically leads to a decrease in the number of individuals supporting opinion $\A$.
Once $\R_{0}^{\A}$ exceeds $1$, the number of individuals supporting opinion $\A$ can increase without relying on opinion $\B$.
To reduce the fraction of supporters of opinion $\B$, $\rho_{\B}$, to zero, having a large $\R_{0}^{\B \to \A}$ is insufficient; $\R_{0}^{\A}$ must exceed $1$.

Fig.~\ref{Fig_phase_diagram_3} shows the case where changing opinion is easy for both opinions ($\R_{0}^{\B \to \A}, \R_{0}^{\A \to \B} > 1$).
In this case, the $\C$ phase exists.
For both opinions, if supporters eagerly propagate the rumor to recruit new supporters ($\R_{0}^{\B \to \A}, \R_{0}^{\A \to \B} > 1$) 
but neglect to maintain existing supporters ($\R_{0}^{\A}, \R_{0}^{\B} < 1$), 
individuals oscillate between opinions $\A$ and $\B$, allowing the two rumors to coexist.

\subsection{Analysis via Perturbation Method}

In reality, individuals rarely change their opinions.
Therefore, we investigate the regime where $\R_{0}^{\B \to \A}, \R_{0}^{\A \to \B} \ll 1$ in detail.
We solve the system of ordinary differential equations approximately using perturbation theory.

\subsubsection{Case where no opinion change occurs}

We consider the case where no opinion change occurs (i.e., the transmission rates $\beta_{\A \to \B} = \beta_{\B \to \A} = 0$).
In this case, since $\dot{\rho}_{\A} = \dot{S}_{\A} + \dot{I}_{\A} = 0$, the fraction of individuals supporting opinion $\A$, $\rho_{\A}$, remains constant from its initial state.
Substituting $S_\A = \rho_\A - I_\A$, we obtain
\begin{align}
    \dot{I}_{\A} 
    &= - \beta_{\A \to \A} I_{\A} \left( I_{\A} - \rho_{\A} + \frac{1}{\R_{0}^{\A}} \right).
    \label{Eq_I_A_2}
\end{align}
Thus, $(S_{\A}, I_{\A}) = (S_{\A}^{\ss}, I_{\A}'^{\ss}(\rho_{\A}))$ given by
\begin{align}
    I_{\A}'^{\ss} (\rho_{\A}) \equiv \rho_{\A} - \frac{1}{\R_{0}^{\A}},\qquad
    S_{\A}^{\ss} \equiv \frac{1}{\R_{0}^{\A}}
\end{align}
represents the steady state.
Note that if $\rho_{r} = 1$, the above expression becomes identical to the steady state of the standard SIS model.

From Eq.~\eqref{Eq_I_A_2}, when $\R_{0}^{\A} \rho_{\A} < 1$, $I_{\A}$ decreases, and no pandemic occurs.
When $\R_{0}^{\A} \rho_{\A} > 1$, we have $0 < I_{\A}'^{\ss} < 1$, and the steady state $I_{\A} = I_{\A}'^{\ss}$ becomes stable.
That is, a transcritical bifurcation\cite{Strogatz2001} occurs at $\R_{0}^{\A} \rho_{\A} = 1$, where the stability of the steady states $I_{\A} = 0$ and $I_{\A} = I_{\A}'^{\ss}$ is exchanged.

\subsubsection{Case where opinion change is rare}

We consider the case where opinion change is rare (i.e., $\beta_{\A \to \B}$, $\beta_{\B \to \A} \ll \beta_{\A \to \A}$, $\beta_{\B \to \B}$, $\gamma_{\A}$, $\gamma_{\B} $).
We solve the ordinary differential equations using the singular perturbation method\cite{Takiguchi2024}, treating $\beta_{\A \to \B}$ and $\beta_{\B \to \A}$ as small parameters (see \ref{Sec_perturbation} for a more rigorous discussion).

First, we consider the expansion up to the zeroth order of the small parameters.
Ignoring transitions between $\A$ and $\B$ yields the equations for the case where no opinion change occurs.
For opinion $r \in \{\A,\B\}$, the system relaxes to the steady state $(S_r, I_r) = (S_r^{\ss}, I_r'^{\ss})$ if $\R_{0}^{r} \rho_r > 1$, and to $(S_r, I_r) = (\rho_r, 0)$ otherwise.

Next, we consider the expansion up to the first order of the small parameters.
The time evolution of the fraction of individuals supporting opinion $\A$, $\rho_{\A} = S_\A + I_\A$, is given by
\begin{align}
\dot{\rho}_{\A}
&= - \beta_{\A \to \B} S_{\A} I_{\B} + \beta_{\B \to \A} S_{\B} I_{\A}.
\label{Eq_rho_A_1}
\end{align}
We substitute $S_{r}(t)$ and $I_{r}(t)$ from the case where no opinion change occurs into this equation.
Note that no terms of the zeroth order of the small parameters appear on the right-hand side of the above equation.
The time evolution of $S_{r}$ and $I_{r}$ driven by the zeroth-order terms is sufficiently faster than that of $\rho_{\A}$.
This implies that the time scale of rumor propagation can be separated from the time scale of opinion change.
When calculating the time evolution of $\rho_{\A}$, we can assume that $S_{r}$ and $I_{r}$ have already relaxed.
Therefore, substituting the steady states of the case where no opinion change occurs into Eq.~\eqref{Eq_rho_A_1} is sufficient.
We perform the calculation separately depending on whether the system relaxes to the steady state $(S_r, I_r) = (S_r^{\ss}, I_r'^{\ss})$ or $(\rho_r, 0)$.

\subsubsection*{(i) Case where $\rho_{\A} > 1 / \R_{0}^{\A} $ and $\rho_{\B} > 1 / \R_{0}^{\B}$}

\begin{figure}[hbtp]
    \centering
    \includegraphics[width=0.6\linewidth]{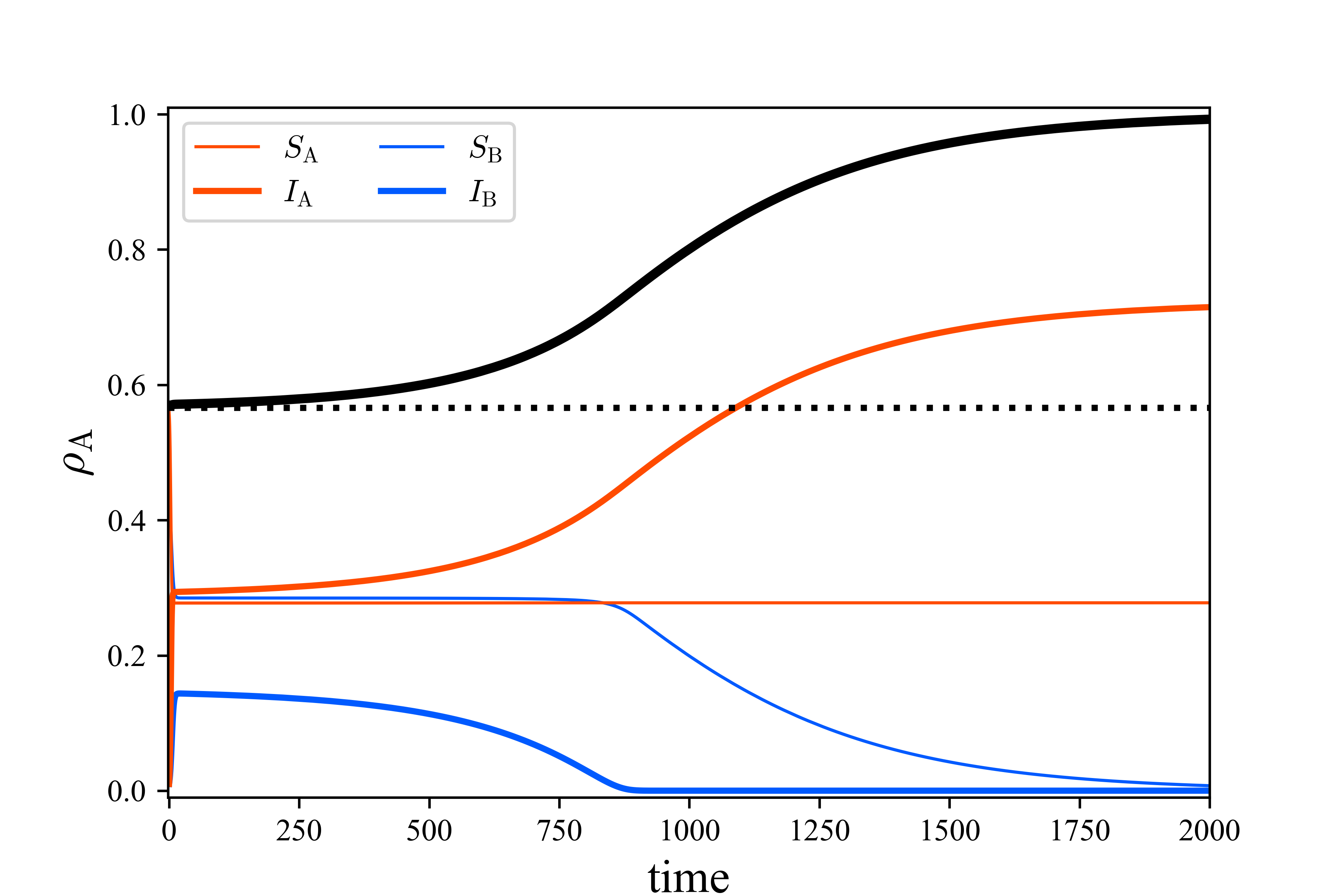}
    \includegraphics[width=0.6\linewidth]{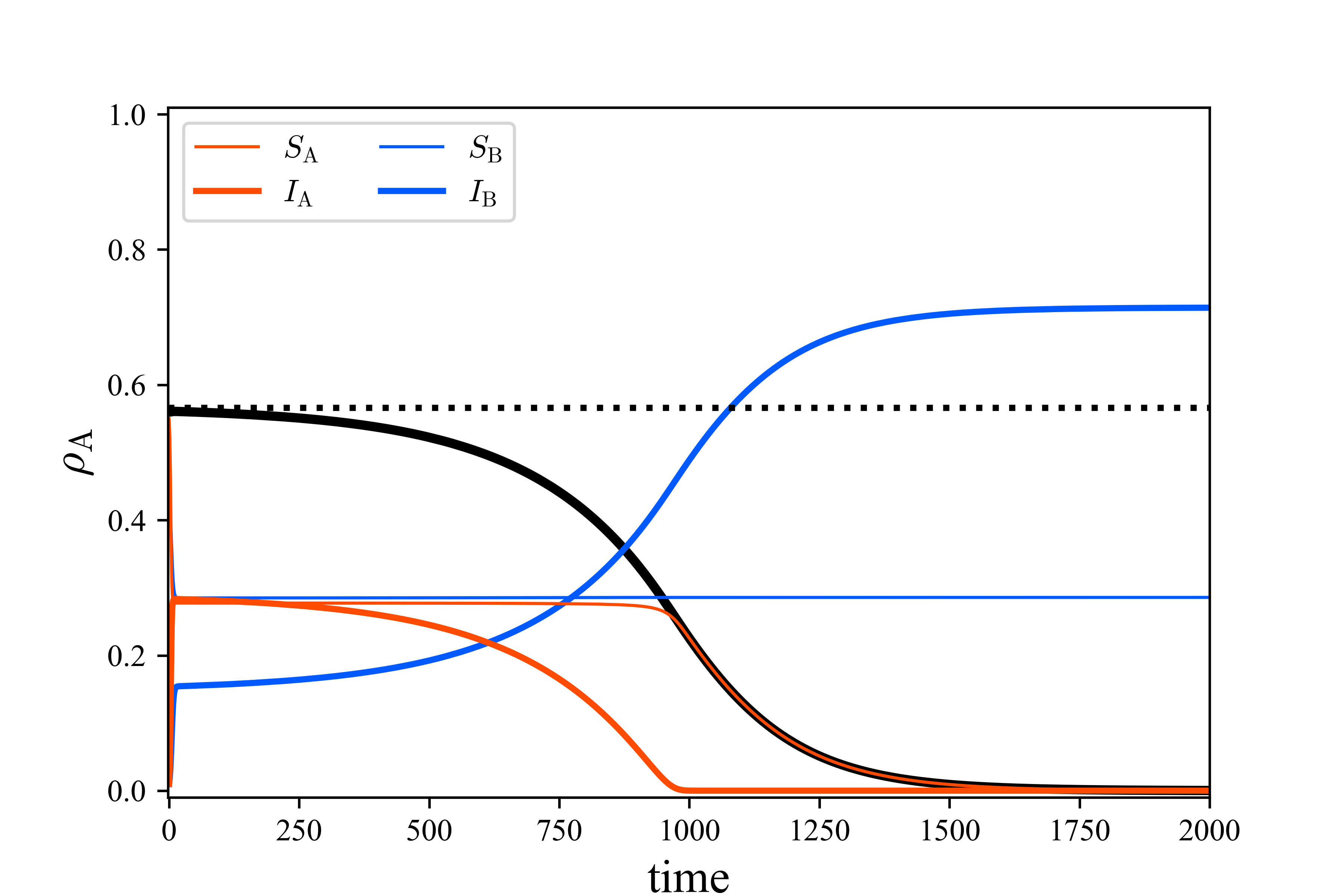}
    \caption{
    Numerical results for the case where opinion change is rare.
    The horizontal axis represents time, and the vertical axis represents the fraction of each state.
    The red and blue lines represent the fractions of individuals supporting opinions $\A$ and $\B$, respectively. 
    The black solid line represents $\rho_{\A}$, and the black dotted line represents the unstable steady state $\rho_{\A}^{\ss}$.
    Parameters: $\gamma_{\A} = \gamma_{\B} = 1.0, \beta_{\A \to \A} = 3.6, \beta_{\B \to \B} = 3.5, \beta_{\A \to \B} = 0.01, \beta_{\B \to \A} = 0.005$.
    In this case, $\rho_{\A}^{\ss} \approx 0.5660\ldots$.
    The initial conditions are $S_{\A} = 0.56, I_{\A} = 0.01, S_{\B} = 0.42, I_{\B} = 0.01$ for the upper panel, 
    and $S_{\A} = 0.55, I_{\A} = 0.01, S_{\B} = 0.43, I_{\B} = 0.01$ for the lower panel.
    \label{Fig_ODE_rho_A_ss}
    }
\end{figure}

In this case, the system first relaxes to $(S_{\A}, I_{\A}, S_{\B}, I_{\B}) = (S_{\A}^{\ss}, I_{\A}'^{\ss}(\rho_{\A}), S_{\B}^{\ss}, I_{\B}'^{\ss}(\rho_{\B}))$.
Substituting this steady state into Eq.~\eqref{Eq_rho_A_1} and calculating yields
\begin{align}
    \dot{\rho}_{\A} 
    &= - \beta_{\A \to \B} S_{\A}^{\ss} I_{\B}'^{\ss} + \beta_{\B \to \A} S_{\B}^{\ss} I_{\A}'^{\ss}\\
    &=
    \left( \frac{\beta_{\A \to \B} \beta_{\B \to \B}\gamma_{\A} + \beta_{\B \to \A} \beta_{\A \to \A} \gamma_{\B}}{\beta_{\A \to \A}\beta_{\B \to \B}} \right)\nonumber\\
    &\qquad \times
    \left( \rho_{\A} - \frac{\beta_{\A \to \B} \beta_{\B \to \B}\gamma_{\A} + (\beta_{\B \to \A} - \beta_{\A \to \B}) \gamma_{\A} \gamma_{\B}}{\beta_{\A \to \B} \beta_{\B \to \B}\gamma_{\A} + \beta_{\B \to \A} \beta_{\A \to \A} \gamma_{\B}} \right)\\
    &\equiv \alpha (\rho_{\A} - \rho_{\A}^{\ss}).
    \label{Eq_rho_A_2}
\end{align}
Note that $0 \leq \rho_{\A}^{\ss} \leq 1$ is satisfied.
Therefore, if $\rho_{\A} > \rho_{\A}^{\ss}$, $\rho_{\A}$ increases, and if $\rho_{\A} < \rho_{\A}^{\ss}$, it decreases.
The state $\rho_{\A} = \rho_{\A}^{\ss}$ is an unstable steady state.
The numerical results of the time evolution of $\rho_{\A}$ are shown in Fig.~\ref{Fig_ODE_rho_A_ss}.
The upper panel corresponds to the case where $\rho_{\A} > \rho_{\A}^{\ss}$ initially, while the lower panel corresponds to the case where $\rho_{\A} < \rho_{\A}^{\ss}$.
The threshold determining which opinion survives is given by $\rho_{\A}^{\ss}$.

The fact that $S_{\A}^{\ss}$ is independent of $\rho_{\A}$ plays a key role in the process where $\rho_{\A}$ diverges from the steady state $\rho_{\A}^{\ss}$.
Individuals in state $\I_{\A}$ can recruit new supporters for opinion $\A$, whereas those in state $\S_{\A}$ cannot.
Furthermore, while individuals in state $\S_{\A}$ may change their opinion, those in state $\I_{\A}$ never do.
Thus, the larger the proportion of $S_{\A}$ within $\rho_{\A}$, the more vulnerable opinion $\A$ becomes to opinion $\B$.
When the system maintains the state $(S_{\A}^{\ss}, I_{\A}'^{\ss}(\rho_{\A}))$, since $S_{\A}^{\ss}$ is constant regardless of $\rho_{\A}$, the smaller $\rho_{\A}$ is, the larger the proportion of $S_{\A}$ within $\rho_{\A}$ becomes.
Consequently, opinion $\A$ becomes more vulnerable to opinion $\B$, causing $\rho_{\A}$ to decrease further. This, in turn, increases the share of $S_{\A}$, making opinion $\A$ even more vulnerable, leading to a runaway decrease in $\rho_{\A}$.
Conversely, in the case where $\rho_{\A} > \rho_{\A}^{\ss}$, the proportion of $S_{\A}$ decreases as $\rho_{\A}$ increases.
In this manner, the system evolves away from $\rho_{\A}^{\ss}$.
In the parameter region where the rate of opinion change is sufficiently small, coexistence of rumors does not occur because the rumor spreads more easily as the number of its supporters increases.

\begin{figure}[hbtp]
    \centering
    \includegraphics[width=0.6\linewidth]{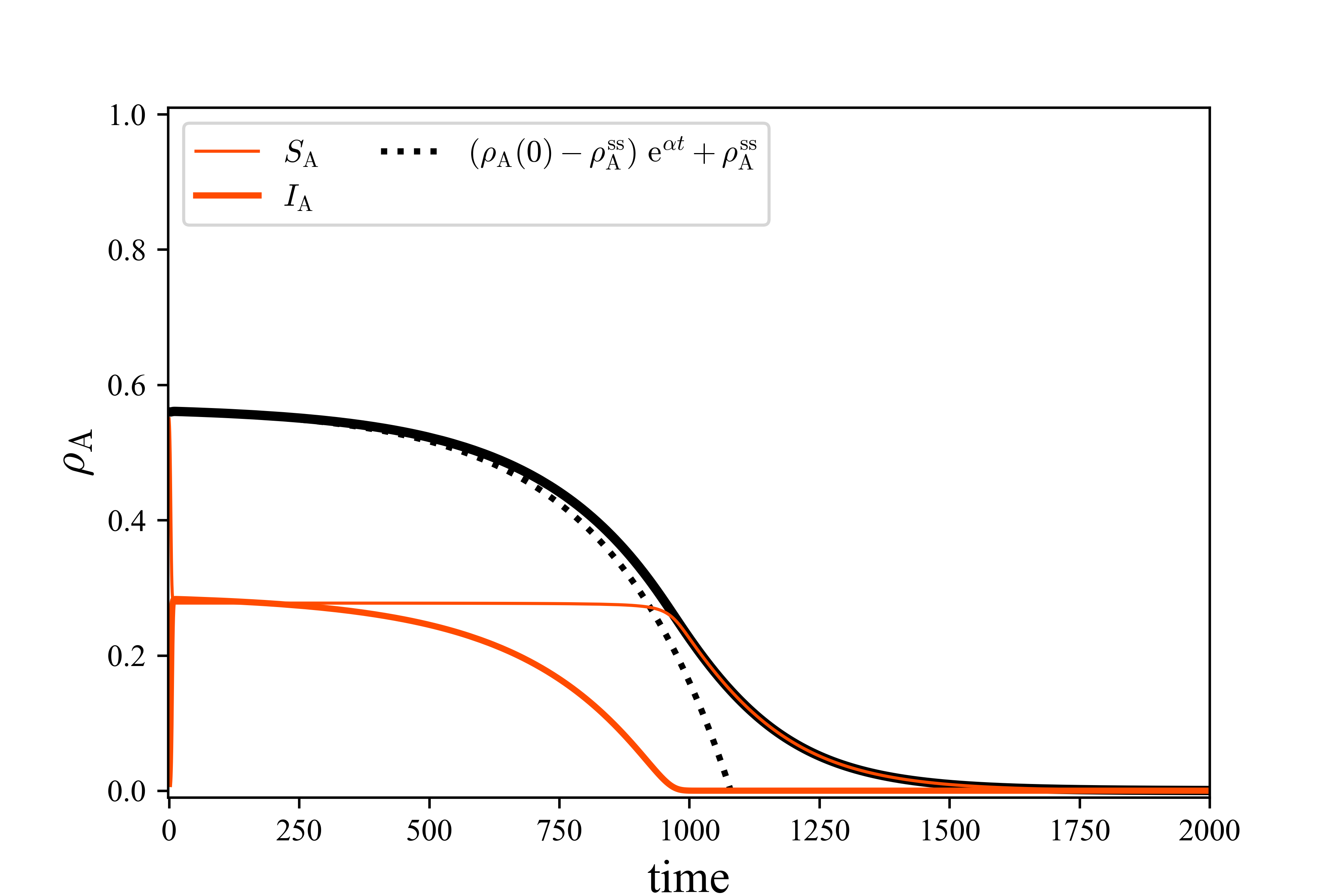}
    \caption{
        Comparison between $\rho_{\A}(t)$ in Fig.~\ref{Fig_ODE_rho_A_ss} and its approximation.
        The solid line represents the numerical result, and the dotted line represents the approximation given by $\rho_{\A}(t) = (\rho_{\A}(0) - \rho_{\A}^{\ss}) \e^{\alpha t } + \rho_{\A}^{\ss}$.
        \label{Fig_ODE_approximation_1}
    }
\end{figure}

The solution to Eq.~\eqref{Eq_rho_A_2} is given by
\begin{align}
\rho_{\A}(t) &= (\rho_{\A}(0) - \rho_{\A}^{\ss}) \e^{\alpha t } + \rho_{\A}^{\ss}.
\label{Eq_rho_A_3}
\end{align}
The comparison between this analytical approximation and the numerical result is shown in Fig.~\ref{Fig_ODE_approximation_1}.
Note that time $t=0$ corresponds not to the initial state of the system, but to the time after it has relaxed to $(S_{\A}, I_{\A}, S_{\B}, I_{\B}) = (S_{\A}^{\ss}, I_{\A}'^{\ss}(\rho_{\A}), S_{\B}^{\ss}, I_{\B}'^{\ss}(\rho_{\B}))$.
In deriving Eq.~\eqref{Eq_rho_A_3}, we assumed that the conditions $\rho_{\A} > 1 / \R_{0}^{\A} $ and $\rho_{\B} > 1 / \R_{0}^{\B}$ hold, 
and thus the system maintains the state $(S_{\A}, I_{\A}, S_{\B}, I_{\B}) = (S_{\A}^{\ss}, I_{\A}'^{\ss}(\rho_{\A}), S_{\B}^{\ss}, I_{\B}'^{\ss}(\rho_{\B}))$.
However, for example, as $\rho_{\A}$ decreases according to Eq.~\eqref{Eq_rho_A_3}, the condition $\rho_{\A} > 1 / \R_{0}^{\A}$ eventually ceases to hold.
Fig.~\ref{Fig_ODE_approximation_1} shows that the approximation breaks down after the time when $\rho_{\A} \approx 1 / \R_{0}^{\A}$, i.e., $I_{\A} \approx 0$.
For $\rho_{\A} < 1 / \R_{0}^{\A} $, we should assume the system maintains $(S_{\A}, I_{\A}, S_{\B}, I_{\B}) = (\rho_{\A}, 0, S_{\B}^{\ss}, I_{\B}'^{\ss}(\rho_{\B}))$.
Conversely, if $\rho_{\A}$ increases, the condition $\rho_{\B} > 1 / \R_{0}^{\B}$ eventually ceases to hold.
In any case, the system dynamics eventually cease to follow Eq.~\eqref{Eq_rho_A_3}.

\subsubsection*{(ii) Case where $\rho_{\A} \leq 1 / \R_{0}^{\A} $ and $\rho_{\B} > 1 / \R_{0}^{\B}$}

\begin{figure}[hbtp]
    \centering
    \includegraphics[width=0.6\linewidth]{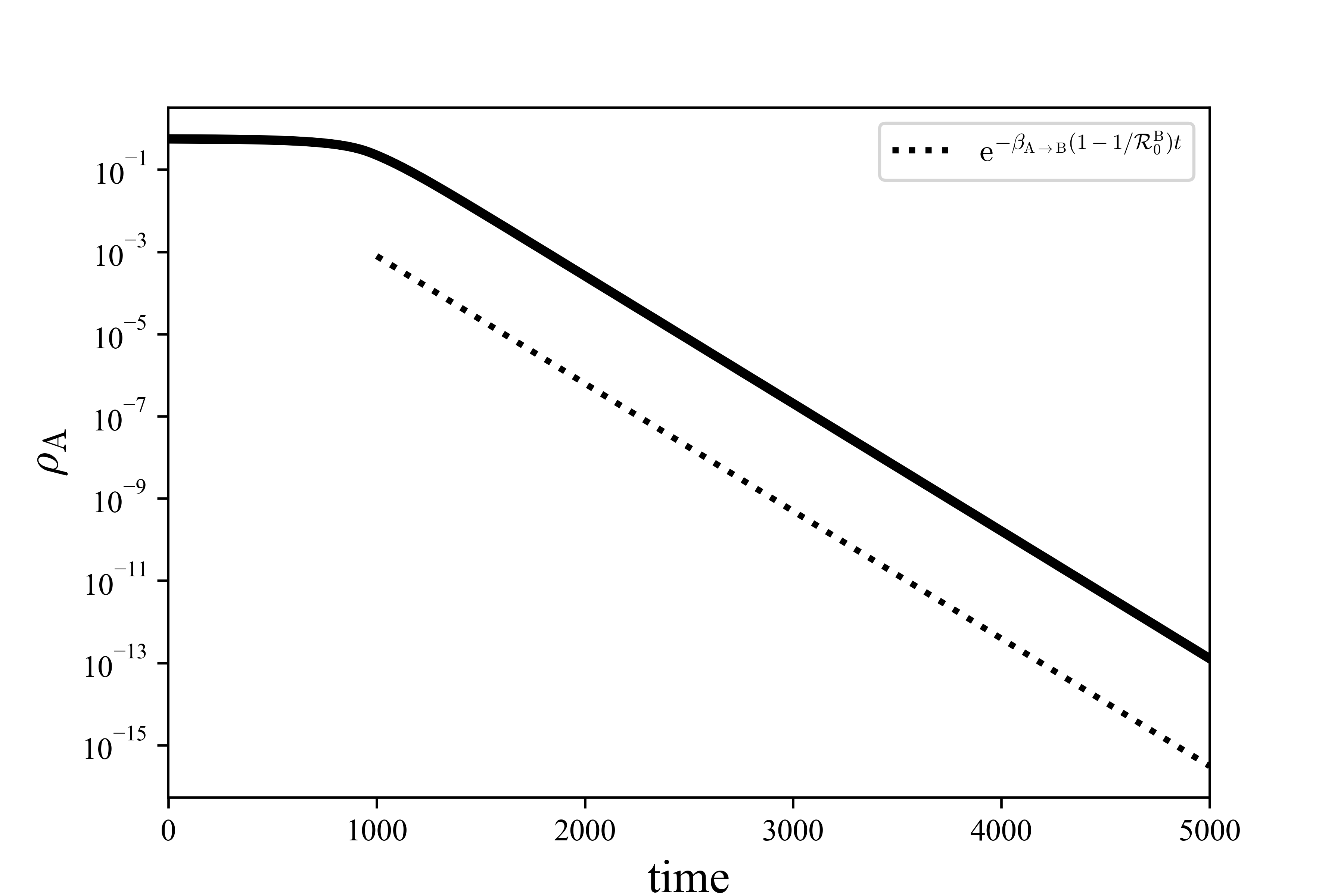}
    \caption{
        Comparison between $\rho_{\A}(t)$ in Fig.~\ref{Fig_ODE_rho_A_ss} and its approximation.
        The solid line represents the numerical result, and the dotted line represents the approximation given by $\rho_{\A}(t) \sim \e^{- \beta_{\A \to \B} (1 - 1/\R_{0}^{\B}) t}$.
        \label{Fig_ODE_approximation_2}
    }
\end{figure}

In this case, the system maintains the state $(S_{\A}, I_{\A}, S_{\B}, I_{\B}) = (\rho_{\A}, 0, S_{\B}^{\ss}, I_{\B}'^{\ss}(\rho_{\B}))$.
Substituting this steady state into Eq.~\eqref{Eq_rho_A_1} and solving this equation yields
\begin{align}
    \rho_{\A}(t)
    &=
    \rho_{\A}(0) \frac{1 - 1/\R_{0}^{\B}}{\rho_{\A}(0) + (1 - 1/\R_{0}^{\B} - \rho_{\A}(0)) \e^{\beta_{\A \to \B} (1 - 1/\R_{0}^{\B}) t} }.
    \label{Eq_rho_A_4}
\end{align}
Note that time $t=0$ corresponds not to the initial state of the system, but to the time after it has relaxed to $(S_{\A}, I_{\A}, S_{\B}, I_{\B}) = (\rho_{\A}, 0, S_{\B}^{\ss}, I_{\B}'^{\ss}(\rho_{\B}))$.
Since $\rho_{\B} > 1 / \R_{0}^{\B}$, the inequalities $1 - 1/\R_{0}^{\B} - \rho_{\A}(0) > 0 $ and $1 - 1/\R_{0}^{\B} > 0$ are satisfied.
For $t \gg 1$, since $\rho_{\A} \sim \e^{- \beta_{\A \to \B} (1 - 1/\R_{0}^{\B}) t} $, $\rho_{\A}$ approaches zero exponentially.
The comparison between the analytical approximation and the numerical result is shown in Fig.~\ref{Fig_ODE_approximation_2}.
The relaxation rate obtained by the perturbation method agrees with the numerical result.

\subsubsection*{(iii) Case where $\rho_{\A} > 1 / \R_{0}^{\A} $ and $\rho_{\B} \leq 1 / \R_{0}^{\B}$}

In this case, the system maintains the state $(S_{\A}, I_{\A}, S_{\B}, I_{\B}) = (S_{\A}^{\ss}, I_{\A}'^{\ss}(\rho_{\A}), \rho_{\B}, 0)$.
Performing a calculation similar to (ii) yields
\begin{align}
    \rho_{\A}(t)
    &=
    1 - (1 - \rho_{\A}(0)) \frac{1 - 1/\R_{0}^{\A}}{(1 - \rho_{\A}(0)) + ( \rho_{\A}(0) - 1/\R_{0}^{\A} ) \e^{\beta_{\B \to \A} (1 - 1/\R_{0}^{\A}) t}}.
\end{align}
For $t \gg 1$, since $1 - \rho_{\A} \sim \e^{- \beta_{\B \to \A} (1 - 1/\R_{0}^{\A}) t} $, $\rho_{\A}$ approaches $1$ exponentially.

\subsubsection*{(iv) Case where $\rho_{\A} \leq 1 / \R_{0}^{\A} $ and $\rho_{\B} \leq 1 / \R_{0}^{\B}$}

In this case, the system first relaxes to $(S_{\A}, I_{\A}, S_{\B}, I_{\B}) = (\rho_{\A}, 0, \rho_{\B}, 0)$.
Since this steady state is an absorbing state, the system does not evolve from this state due to the small parameters.
Indeed, substituting this steady state into Eq.~\eqref{Eq_rho_A_1} yields $\dot{\rho}_{\A} = 0$.

\subsubsection*{Summary}

Summarizing the results from (i) to (iv), the time evolution of $\rho_{\A}$ is given by the following equations:
\begin{align}
    \rho_{\A}(t) &= (\rho_{\A}(0) - \rho_{\A}^{\ss}) \e^{\alpha t } + \rho_{\A}^{\ss} && \left(1/\R_{0}^{\A} < \rho_{\A}(t) < 1 - 1/\R_{0}^{\B} \right),\\
    \rho_{\A}(t) &\sim \e^{- \beta_{\A \to \B} (1 - 1/\R_{0}^{\B}) t} && \left(\rho_{\A}(t) \leq 1/\R_{0}^{\A}, 1 - 1/\R_{0}^{\B} \right),\\
    1 - \rho_{\A}(t) &\sim \e^{- \beta_{\B \to \A} (1 - 1/\R_{0}^{\A}) t} && \left(1/\R_{0}^{\A}, 1 - 1/\R_{0}^{\B} \leq \rho_{\A}(t)\right),\\
    \rho_{\A}(t) &= \const && \left(1 - 1/\R_{0}^{\B} \leq \rho_{\A}(t) \leq 1/\R_{0}^{\A}\right).
\end{align}
Accelerating away from the threshold $\rho_{\A} = \rho_{\A}^{\ss}$, the system asymptotically approaches either $\rho_{\A} = 0$ or $1$.
Ultimately, only one of the opinions $\A$ or $\B$ survives, and the outcome is determined by the initial state.
Even if the parameters favor the spread of the rumor $\A$ (i.e., $\R_{0}^{\A} > \R_{0}^{\B}$ and $\R_{0}^{\B \to \A} > \R_{0}^{\A \to \B}$), 
if the fraction of supporters of opinion $\A$ does not exceed the threshold $\rho_{\A}^{\ss}$, 
this opinion will be driven out by opinion $\B$.
Although this model does not explicitly incorporate the majority conformity effect, 
the property that an opinion becomes stronger as the number of its supporters increases emerges spontaneously.

The approximate calculations performed here become more accurate as the opinion change rates $\beta_{\A \to \B}$ and $\beta_{\B \to \A}$ approach zero.
Even when the opinion change rates cannot be regarded as small parameters, the behavior can be described qualitatively correctly provided that $\beta_{\A \to \B}, \beta_{\B \to \A} < \beta_{\A \to \A}, \beta_{\B \to \B}, \gamma_{\A}, \gamma_{\B} $.

\section{Conclusions and Discussion}

We proposed the SIS competition model as a simple model capable of describing both rumor propagation and opinion dynamics.
In this model, steady states and their local stability can be determined analytically.
In addition, when opinion change is rare, the time evolution of the fraction of supporters for each opinion can be calculated analytically using perturbation methods.
From the linear stability analysis, we found that there are two conditions for only rumor $r$ to survive: (i) it can cause a pandemic on its own ($\R_{0}^{r} > 1$), and (ii) it is more transmissible than the other rumor $r'$ in a situation where only silent supporters of opinion $r$ exist ($\R_{0}^{r} > \R_{0}^{r'}$).
When both rumors satisfy these conditions, which rumor survives is determined by the fraction of supporters at the initial time.
In the case where opinion change is rare, this threshold can be determined analytically.

The SIS competition model possesses four steady states: a state where neither rumor spreads, states where only rumor $\A$ or $\B$ spreads, and a state where both rumors coexist.
Although this result appears similar to the Lotka-Volterra competition model\cite{Murray2007}, the mechanism of coexistence is entirely different.
In the Lotka-Volterra competition model, different species are in an antagonistic relationship, and coexistence occurs when the parameters representing interspecies interaction are sufficiently small.
In contrast, in our model, different opinions are not necessarily antagonistic, and coexistence occurs when the parameters representing the interaction between opinions ($\beta_{\B \to \A}$, $\beta_{\A \to \B}$) are sufficiently large.
The mechanism of coexistence is to maintain rumor diffusion by causing supporters of other opinions to change their minds.
If corrective information is trivial to those who know the truth but shocking to those who believe fake news, yet insufficient to eradicate the fake news, these incompatible opinions can coexist. 
Despite fake news and corrective information being incompatible, people believing fake news paradoxically aid the spread of corrective information.
Since this corrective information cannot survive on its own, once the people spreading fake news disappear, the spread of corrective information also stops. 
Our model can describe the nontrivial but realistic phenomenon where a smaller infection rate actually allows a rumor to spread. 
Note that since there are cases in reality where opinions coexist even when opinion change is rare, we do not consider that all coexistence phenomena can be explained by this mechanism.

Our results offer the following two implications.
First, opinion $\B$ cannot reduce the supporters of opinion $\A$ to zero unless $\R_{0}^{\B}$ exceeds $1$, no matter how large $\R_{0}^{\A \to \B}$ is.
Letting opinion $\A$ be fake news and opinion $\B$ be corrective information, this implies that to eliminate believers of fake news, the corrective information must be interesting enough to survive on its own.
Second, even though this model does not explicitly incorporate the majority conformity effect, in the case where opinion change is rare, the property that a rumor spreads more easily as it has more supporters emerges spontaneously.
Therefore, when an existing opinion is widely accepted, it is difficult for a conflicting opinion to overturn it, even without majority conformity.

Because our proposed model is simple, it is easy to extend. 
Possible directions for future research include adding a state of supporting neither opinion, introducing terms corresponding to mass media, or incorporating social networks. 
When introducing a network with community structure, in the case where opinion change is rare (and thus not in the coexistence phase), 
a phenomenon can occur where one opinion dominates one community and the other dominates another, leading to the coexistence of multiple opinions in the population as a whole. 
We plan to report on this phenomenon in a future paper.

\section*{Acknowledgments}

This work was supported by JST SPRING, Grant Number JPMJSP2119.

\appendix

\section{Linear Stability Analysis of the Coexistence Steady State}
\label{Sec_Hurwitz}

We derive the conditions under which the steady state where both rumors coexist becomes locally stable.
From the characteristic equation of the Jacobian matrix, one eigenvalue is $\lambda = 0$, and the remaining three eigenvalues are the roots of the following cubic equation:
\begin{align}
    0
    =& \
    \lambda^3
    + [(\beta_{\A \to \A} + \beta_{\B \to \A} ) I_{\A}^{*\ss} + (\beta_{\B \to \B} + \beta_{\A \to \B} ) I_{\B}^{*\ss}] \lambda^2\nonumber\\
    &
    + 
    \left[ 
        (\beta_{\B \to \A} - \beta_{\A \to \A}) \beta_{\B \to \A} S_{\B}^{*\ss} I_{\A}^{*\ss}
        + (\beta_{\A \to \B} - \beta_{\B \to \B}) \beta_{\A \to \B} S_{\A}^{*\ss}  I_{\B}^{*\ss}
    \right.\nonumber\\
    & \qquad
    \left.
        + (\beta_{\A \to \A} I_{\A}^{*\ss} + \beta_{\A \to \B} I_{\B}^{*\ss}) (\beta_{\B \to \B} I_{\B}^{*\ss} + \beta_{\B \to \A} I_{\A}^{*\ss})
    \right] \lambda\nonumber\\
    &
    + 
    (\beta_{\B \to \A} \beta_{\A \to \B} - \beta_{\A \to \A} \beta_{\B \to \B} ) (\beta_{\A \to \B} S_{\A}^{*\ss} + \beta_{\B \to \A} S_{\B}^{*\ss} ) I_{\A}^{*\ss} I_{\B}^{*\ss}.
    \label{Eq_lambda_equation}
\end{align}
Although the eigenvalues can be theoretically obtained using the general formula for cubic equations, the calculations are prohibitively complex.
Therefore, we discuss the stability without explicitly solving for the eigenvalues.
Using the Hurwitz stability criterion\cite{Murray2007}, 
the necessary and sufficient conditions for the real parts of all roots of the cubic equation $\lambda^3 + a_1 \lambda^2 + a_2 \lambda + a_3 = 0$ to be negative are $a_1 > 0$, $a_3 > 0$, and $a_1 a_2 - a_3 > 0$.
We determine the parameter range in Eq.~\eqref{Eq_lambda_equation} that satisfies these conditions.
It is sufficient to consider only the parameter region where the steady state is physically meaningful, i.e., $0 \leq S_{\A}^{*\ss}, I_{\A}^{*\ss}, S_{\B}^{*\ss}, I_{\B}^{*\ss} \leq 1$.
The condition $a_1 > 0$ is always satisfied.
The condition $a_3 > 0$ holds when $\beta_{\B \to \A} \beta_{\A \to \B} > \beta_{\A \to \A} \beta_{\B \to \B}$, that is, $\R_{0}^{\B \to \A} \R_{0}^{\A \to \B} > \R_{0}^{\A} \R_{0}^{\B}$.
Recall that the conditions $0 \leq S_{\A}^{*\ss}$ and $0 \leq S_{\B}^{*\ss}$ are satisfied when the signs of $\R_{0}^{\A \to \B} - \R_{0}^{\A}$ and $\R_{0}^{\B \to \A} - \R_{0}^{\B}$ match.
Therefore, for the stable steady state, it is necessary that $\R_{0}^{\A \to \B} > \R_{0}^{\A}$ and $\R_{0}^{\B \to \A} > \R_{0}^{\B}$.
Finally, we consider $\Delta \equiv a_1 a_2 - a_3$.
\begin{align}
    \Delta
    \geq&
    + [(\beta_{\A \to \A} + \beta_{\B \to \A} ) I_{\A}^{*\ss} + (\beta_{\B \to \B} + \beta_{\A \to \B} ) I_{\B}^{*\ss}] (\beta_{\B \to \A} - \beta_{\A \to \A}) \beta_{\B \to \A} S_{\B}^{*\ss} I_{\A}^{*\ss}\nonumber\\
    & + [(\beta_{\A \to \A} + \beta_{\B \to \A} ) I_{\A}^{*\ss} + (\beta_{\B \to \B} + \beta_{\A \to \B} ) I_{\B}^{*\ss}] (\beta_{\A \to \B} - \beta_{\B \to \B}) \beta_{\A \to \B} S_{\A}^{*\ss}  I_{\B}^{*\ss}\nonumber\\
    & - (\beta_{\B \to \A} \beta_{\A \to \B} - \beta_{\A \to \A} \beta_{\B \to \B} ) (\beta_{\A \to \B} S_{\A}^{*\ss} + \beta_{\B \to \A} S_{\B}^{*\ss} ) I_{\A}^{*\ss} I_{\B}^{*\ss}\\
    =&
    + (\beta_{\A \to \A} + \beta_{\B \to \A} ) (\beta_{\B \to \A} - \beta_{\A \to \A}) \beta_{\B \to \A} S_{\B}^{*\ss} (I_{\A}^{*\ss})^2 \nonumber\\
    & + (\beta_{\B \to \B} \beta_{\B \to \A} - \beta_{\A \to \A} \beta_{\A \to \B}) \beta_{\B \to \A} S_{\B}^{*\ss} I_{\A}^{*\ss} I_{\B}^{*\ss} \nonumber\\
    & + (\beta_{\A \to \A} \beta_{\A \to \B} - \beta_{\B \to \B} \beta_{\B \to \A}) \beta_{\A \to \B} S_{\A}^{*\ss} I_{\A}^{*\ss} I_{\B}^{*\ss} \nonumber\\
    & + (\beta_{\B \to \B} + \beta_{\A \to \B} ) (\beta_{\A \to \B} - \beta_{\B \to \B}) \beta_{\A \to \B} S_{\A}^{*\ss}  (I_{\B}^{*\ss})^2.
\end{align}
Multiplying the right-hand side of the above equation by $(\beta_{\A \to \B} S_{\A}^{*\ss} + \beta_{\B \to \A} S_{\B}^{*\ss})^2 (1 - S_{\A}^{*\ss} - S_{\B}^{*\ss})^{-2} \ ( \geq 0)$ 
and defining the result as $\Delta'$, we obtain:
\begin{align}
    \Delta'
    =
    & + [\beta_{\B \to \A} (\beta_{\B \to \A} - \beta_{\B \to \B}) + \beta_{\A \to \A} (\beta_{\A \to \B} - \beta_{\A \to \A}) ] \beta_{\B \to \A} S_{\B}^{*\ss} (\beta_{\A \to \B} S_{\A}^{*\ss})^2 \nonumber\\
    & + [\beta_{\A \to \B} (\beta_{\A \to \B} - \beta_{\A \to \A}) + \beta_{\B \to \B} (\beta_{\B \to \A} - \beta_{\B \to \B}) ] \beta_{\A \to \B} S_{\A}^{*\ss}  (\beta_{\B \to \A} S_{\B}^{*\ss})^2.
\end{align}
If $\R_{0}^{\A \to \B} > \R_{0}^{\A}$ and $\R_{0}^{\B \to \A} > \R_{0}^{\B}$ hold, then $\Delta' > 0$ is satisfied.
In conclusion, the steady state where rumors coexist is stable if $\R_{0}^{\A \to \B} > \R_{0}^{\A}$ and $\R_{0}^{\B \to \A} > \R_{0}^{\B}$.

\section{Singular Perturbation Method}
\label{Sec_perturbation}

In this appendix, we provide a more rigorous discussion of the perturbation calculation for the case where opinion change is rare.
We introduce the rescaling of the opinion change parameters as $\beta_{\A \to \B} \to \ep \beta_{\A \to \B}$ and $\beta_{\B \to \A} \to \ep \beta_{\B \to \A}$.
Here, $\ep$ is a small parameter, and we assume that $\beta_{\A \to \B}$ and $\beta_{\B \to \A}$ are of the same order as $\beta_{\A \to \A}$ and $\beta_{\B \to \B}$.
We introduce a sequence of time points $t_0, t_1, \ldots, t_n, \ldots$
Given $t_n$, the next time $t_{n+1}$ is chosen arbitrarily within a range that satisfies
\begin{align}
    \max_{t \in [t_n, t_{n+1}]}
    \left|
        \rho_{\A}(t) - \rho_{\A}(t_n)
    \right|
    = O(\ep).
\end{align}

We expand the fractions of each state $S_r(t)$ and $I_r(t)$ ($r \in \{ \A, \B \}$) for $t \in [t_n, t_{n+1}]$ in powers of $\ep$ as follows:
\begin{align}
    S_r(t) &= S_{r}^{(0;n)}(t) + \ep S_{r}^{(1;n)}(t) + \ep^2 S_{r}^{(2;n)}(t) + \cdots,\label{Eq_S_r_ep}\\
    I_r(t) &= I_{r}^{(0;n)}(t) + \ep I_{r}^{(1;n)}(t) + \ep^2 I_{r}^{(2;n)}(t) + \cdots.\label{Eq_I_r_ep}
\end{align}
Here, the zeroth-order terms satisfy $S_r^{(0;n)}(t_n) = S_r(t_n)$ and $I_r^{(0;n)}(t_n) = I_r(t_n)$, and the terms of order $l$ ($\geq 1$) satisfy $S_r^{(l;n)}(t_n) = 0$ and $I_r^{(l;n)}(t_n) = 0$.
Substituting these expressions into the ordinary differential equations~\eqref{Eq_S_A}-\eqref{Eq_I_B}, the zeroth-order equations are
\begin{align}
    \dot{S}_{r}^{(0;n)}(t) &= - \beta_{r \to r} S_{r}^{(0;n)}(t) I_{r}^{(0;n)}(t) + \gamma_{r} I_{r}^{(0;n)}(t),\label{Eq_S_r_0th}\\  
    \dot{I}_{r}^{(0;n)}(t) &= + \beta_{r \to r} S_{r}^{(0;n)}(t) I_{r}^{(0;n)}(t) - \gamma_{r} I_{r}^{(0;n)}(t),\label{Eq_I_r_0th} 
\end{align}
and the first-order equations are
\begin{align}
    \dot{S}_{r}^{(1;n)}(t) 
    =& - \beta_{r \to r} S_{r}^{(1;n)}(t) I_{r}^{(0;n)}(t) - \beta_{r \to r} S_{r}^{(0;n)}(t) I_{r}^{(1;n)}(t) \nonumber\\
    &- \beta_{r \to r'} S_{r}^{(0;n)}(t) I_{r'}^{(0;n)}(t) + \gamma_{r} I_{r}^{(1;n)}(t)\label{Eq_S_r_1th},\\  
    \dot{I}_{r}^{(1;n)}(t) 
    =& + \beta_{r \to r} S_{r}^{(1;n)}(t) I_{r}^{(0;n)}(t) + \beta_{r \to r} S_{r}^{(0;n)}(t) I_{r}^{(1;n)}(t) \nonumber\\
    &+ \beta_{r' \to r} S_{r'}^{(0;n)}(t) I_{r}^{(0;n)}(t) - \gamma_{r} I_{r}^{(1;n)}(t)\label{Eq_I_r_1th}.
\end{align}
Defining the fraction of supporters of opinion $\A$ as $\rho_{\A}^{(l;n)}(t) \equiv S_{\A}^{(l;n)}(t) + I_{\A}^{(l;n)}(t)$, the time evolution of the zeroth-order term is $\dot{\rho}_{\A}^{(0;n)}(t) = 0$, implying that it takes a constant value $\rho_{\A}^{(0;n)}(t) = \rho_{\A}^{(0;n)}(t_n) \equiv \rho_{\A}^{(0;n)}$.
The first-order time evolution is given by
\begin{align}
    \dot{\rho}_{\A}^{(1;n)}(t) &= - \beta_{\A \to \B} S_{\A}^{(0;n)}(t) I_{\B}^{(0;n)}(t) + \beta_{\B \to \A} S_{\B}^{(0;n)}(t) I_{\A}^{(0;n)}(t),
    \label{Eq_rho_A_1st}
\end{align}
which can be expressed solely in terms of the zeroth-order quantities $S_{r}^{(0;n)}(t)$ and $I_{r}^{(0;n)}(t)$.

The solutions to the zeroth-order ordinary differential equations~\eqref{Eq_S_r_0th} and \eqref{Eq_I_r_0th} are given by
\begin{align}
    I_{r}^{(0;n)}(t) &= \frac{I_{r}'^{\ss}(\rho_{r}^{(0;n)})}{1 - (1 - I_{r}'^{\ss}(\rho_{r}^{(0;n)})/ I_{r}^{(0;n)}(t_n)) \e^{- \beta_{r \to r} I_{r}'^{\ss} (t-t_n)}},\\
    & I_{r}'^{\ss}\left( \rho_{r}^{(0;n)} \right) \equiv \rho_{r}^{(0;n)} - \frac{1}{\R_{0}^{r} },\label{Eq_I_r_0th_tn}\\
    S_{r}^{(0;n)}(t) &= \rho_{r}^{(0;n)} - I_{r}^{(0;n)}(t).
\end{align}
The steady states are $(S_r, I_r) = (1/\R_{0}^{r},\ I_{r}'^{\ss}( \rho_{r}^{(0;n)}))$ and $(\rho_{r}^{(0;n)}, 0)$.

In the perturbation expansion for time $t \in [t_{n-1},t_{n}]$, we assume that the following relation holds for $r \in \{\A,\B\}$:
\begin{align}
    I_{r}^{(0;n-1)}(t_n) &= I_{r}'^{\ss}\left(\rho_{r}^{(0;n-1)}\right) + O(\ep).
    \label{Eq_I_r_0th_tn-1}
\end{align}
That is, we assume that the state of the system at time $t = t_n$ deviates from the steady state $(S_r, I_r) = (1/\R_{0}^{r},\ I_{r}'^{\ss}( \rho_{r}^{(0;n-1)}))$ by only $O(\ep)$.
Under this assumption, we consider the dynamics of the system for time $t \in [t_n,t_{n+1}]$.
Noting that the following equation holds:
\begin{align}
    \rho_{r}^{(0;n)}
    &= \rho_{r}(t_n)
    = \rho_{r}^{(0;n-1)} + \ep \rho_{r}^{(1;n-1)}(t_n) + O(\ep^2),
\end{align}
we obtain $I_r^{(0;n)}(t_n) = I_{r}'^{\ss}(\rho_{r}^{(0;n)}) + O(\ep)$ from Eq.~\eqref{Eq_I_r_0th_tn-1}.
Substituting this into the expression for $I_{r}^{(0;n)}(t)$ in Eq.~\eqref{Eq_I_r_0th_tn} yields $I_{r}^{(0;n)}(t) = I_{r}'^{\ss}(\rho_{r}^{(0;n)}) + O(\ep)$.
We also obtain $S_{r}^{(0;n)}(t) = 1/\R_{0}^{r} + O(\ep)$.
In other words, for time $t \in [t_n,t_{n+1}]$, $(I_{r}^{(0;n)}(t), S_{r}^{(0;n)}(t))$ does not change with time at the zeroth order of $\ep$ and maintains the steady state $(1/\R_{0}^{r},\ I_{r}'^{\ss}( \rho_{r}^{(0;n)}))$.

Substituting this into the time evolution equation for $\rho_{\A}^{(1;n)}$, Eq.~\eqref{Eq_rho_A_1st}, yields
\begin{align}
    \dot{\rho}_{\A}^{(1;n)}(t) 
    &=
    \left( \frac{\beta_{\A \to \B} \beta_{\B \to \B}\gamma_{\A} + \beta_{\B \to \A} \beta_{\A \to \A} \gamma_{\B}}{\beta_{\A \to \A}\beta_{\B \to \B}} \right)\nonumber\\
    &\qquad \times
    \left( \rho_{\A}^{(0)} - \frac{\beta_{\A \to \B} \beta_{\B \to \B}\gamma_{\A} + (\beta_{\B \to \A} - \beta_{\A \to \B}) \gamma_{\A} \gamma_{\B}}{\beta_{\A \to \B} \beta_{\B \to \B}\gamma_{\A} + \beta_{\B \to \A} \beta_{\A \to \A} \gamma_{\B}} \right)
    + O(\ep)\\
    &\equiv \alpha \left(\rho_{\A}^{(0;n)} - \rho_{\A}^{\ss}\right) + O(\ep).
\end{align}
Since $\rho_{\A}(t) = \rho_{\A}^{(0;n)} + \ep \rho_{\A}^{(1;n)}(t) + O(\ep^2)$ for time $t \in [t_{n},t_{n+1}]$, we can calculate from the above equation that:
\begin{align}
    \dot{\rho}_{\A} (t)
    &= \ep \alpha \left(\rho_{\A}(t) - \rho_{\A}^{\ss}\right) + O(\ep^2).
    \label{Eq_rho_A_t}
\end{align}
It can be shown that the above equation continues to hold in a similar manner after time $t = t_{n+1}$.
Note that if Eq.~\eqref{Eq_I_r_0th_tn-1} is assumed, then
\begin{align}
    I_{r}^{(0;n)}(t_{n+1}) &= I_{r}'^{\ss}\left(\rho_{r}^{(0;n)}\right) + O(\ep)
\end{align}
also holds.
That is, if Eq.~\eqref{Eq_I_r_0th_tn-1} holds for $t \in [t_{n-1},t_{n}]$, it can be shown that the corresponding equation holds for subsequent times.
Therefore, letting $t_\i$ be the time when this assumption regarding $I_{r}(t)$ is first satisfied, 
and $t_{\mathrm{f}}$ be the time when $I_{r}'^{\ss}(\rho_{r}^{(0;n)}) = 0$, Eq.~\eqref{Eq_rho_A_t} holds for any time $t \in [t_\i, t_{\mathrm{f}}]$.

Expressions for $\rho_{\A}(t)$ can be obtained in a similar manner for the cases where $\rho_{\A} \leq 1 / \R_{0}^{\A} $ and $\rho_{\B} > 1 / \R_{0}^{\B} $, as well as where $\rho_{\A} > 1 / \R_{0}^{\A} $ and $\rho_{\B} \leq 1 / \R_{0}^{\B} $.
In the case of relaxation to the steady state $(S_r, I_r) = (\rho_{r}^{(0;n)}, 0)$, the dynamics can be calculated by assuming $I_{r}^{(0;n-1)}(t_n) = O(\ep)$.

\begin{figure}[hbtp]
    \centering
    \includegraphics[width=0.6\linewidth]{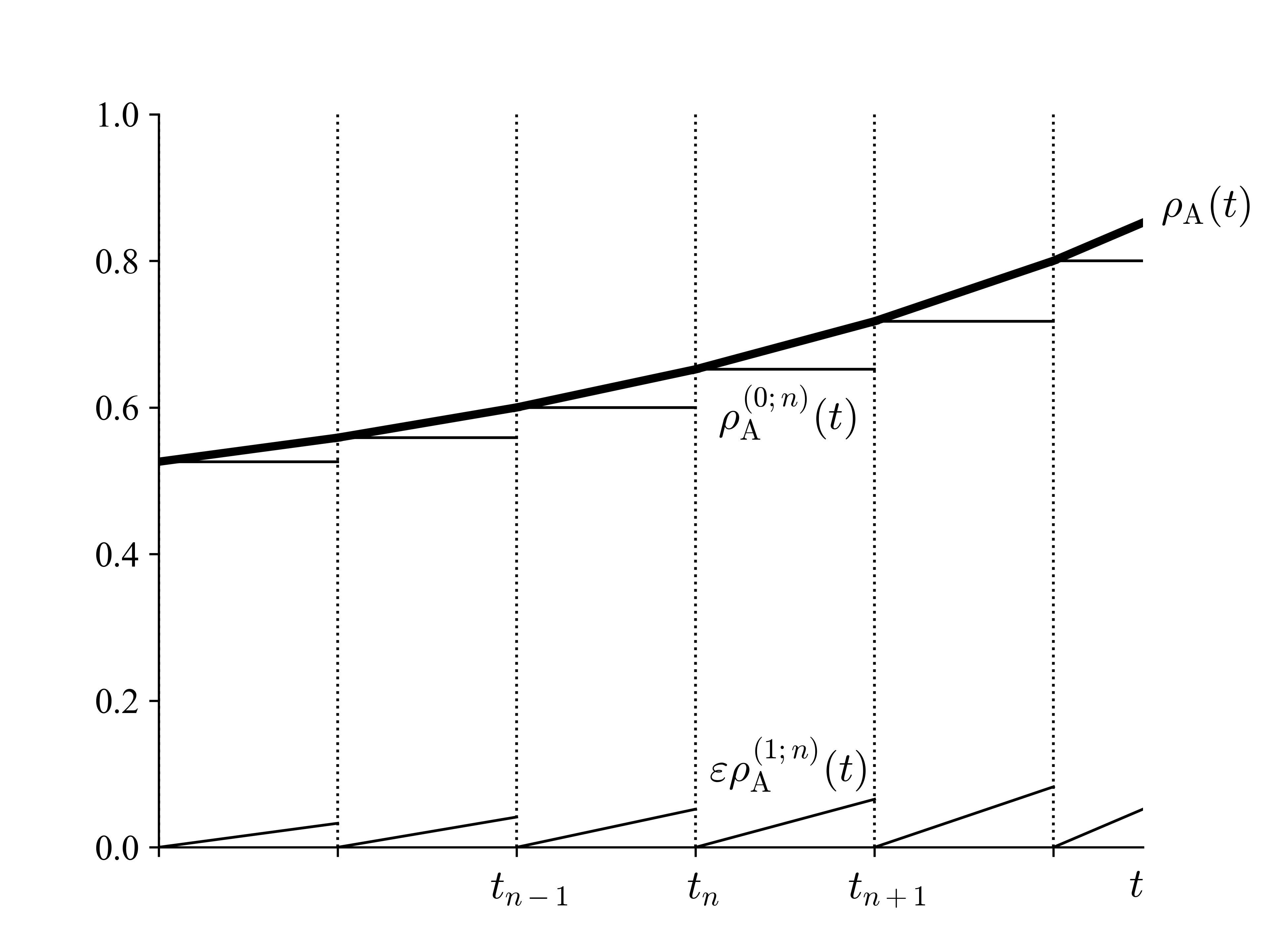}
    \caption{
        Schematic of the singular perturbation method.
        \label{Fig_perturbation}
    }
\end{figure}

Fig.~\ref{Fig_perturbation} illustrates the schematic of the perturbation method used in this appendix.
For the fraction of supporters of opinion $\A$ at time $t$, $\rho_{\A}(t)$, we re-evaluate the perturbation expansion at each time step $t=\ldots, t_{n-1}, t_n, t_{n+1}, \ldots$.
In the time interval $t\in [t_n,t_{n+1}]$, the zeroth-order term of $\rho_{\A}(t)$ is a constant value $\rho_{\A}^{(0;n)}(t) = \rho_{\A}(t_n)$, and the first-order term is a linear function of time $\rho_{\A}^{(1;n)}(t) = \alpha [\rho_{\A}(t_n) - \rho_{\A}^{\ss}] t$.
Thus, by considering terms up to the first order and connecting the graphs at each time step, we obtain a piecewise linear graph.
Taking the limit $\ep \to 0$, the time interval for the perturbation expansion also approaches zero, and this piecewise linear graph becomes a smooth curve.
This corresponds to the solution $\rho_{\A}(t)$ of Eq.~\eqref{Eq_rho_A_t}.

\end{document}